# Designing a Layered Framework to Secure Data via Improved Multi-Stage Lightweight Cryptography in IoT–Cloud Systems

Hojjat Farshadinia[a], Ali Barati[a,*], Hamid Barati[a]

[a] Department of Computer Engineering, Dezful Branch, Islamic Azad University, Dezful, Iran

## Abstract

This paper presents a novel multi-layered hybrid security approach aimed at enhancing lightweight encryption for IoT-Cloud systems. The primary goal is to overcome limitations inherent in conventional solutions such as TPA, Blockchain, ECDSA and ZSS which often fall short in terms of data protection, computational efficiency and scalability. Our proposed method strategically refines and integrates these technologies to address their shortcomings while maximizing their individual strengths. By doing so we create a more reliable and high-performance framework for secure data exchange across heterogeneous environments. The model leverages the combined potential of emerging technologies, particularly Blockchain, IoT and Cloud computing which when effectively coordinated offer significant advancements in security architecture. The proposed framework consists of three core layers: (1) the H.E.EZ Layer which integrates improved versions of Hyperledger Fabric, Enc-Block and a hybrid ECDSA-ZSS scheme to improve encryption speed, scalability and reduce computational cost; (2) the Credential Management Layer independently verifying data integrity and authenticity; and (3) the Time and Auditing Layer designed to reduce traffic overhead and optimize performance across dynamic workloads. Evaluation results highlight that the proposed solution not only strengthens security but also significantly improves execution time, communication efficiency and system responsiveness, offering a robust path forward for next-generation IoT-Cloud infrastructures.

**Keywords:** Hyperledger Fabric blockchain, IoT-Cloud, data security, computation time, overhead

## 1.Introduction

[*] Corresponding author
Email addresses: h.farshadinia@iau.ac.ir (Hojjat Farshadinia), alibarati@iau.ac.ir (Ali Barati), hamid.barati@iau.ac.ir (Hamid Barati)

A third-party auditor refers to an independent external entity or individual responsible for conducting audits to confirm an organization's adherence to established standards or regulations. Their role involves evaluating whether the company’s processes, systems or products comply with specific requirements while maintaining impartiality and objectivity. However, third-party auditors may themselves face certain limitations that could affect the accuracy and trustworthiness of their assessments. These limitations might include insufficient precision, overlooking non-compliance issues, inadequate emphasis on continuous improvement as well as concerns related to auditor qualifications or potential conflicts of interest. Additionally, dependence on external information sources and the risk of receiving misleading data can further compromise the overall quality of the audit process [1–5].

Blockchain represents an innovative form of distributed database technology integrating a variety of emerging IT solutions and playing a crucial role within IoT-Cloud frameworks. Although it may not fully replace third-party auditors, blockchain holds considerable promise to revolutionize auditing practices. By improving transparency and operational efficiency this technology can greatly influence the auditing landscape. It enables automation of audit workflows, minimizes human errors and establishes a trustworthy, cost-efficient audit trail—potentially reducing the need for manual checks and reliance on external auditors [6–10].

ECDSA has emerged as the standard digital signature method for many contemporary blockchain platforms offering solutions to certain security challenges inherent in these systems. By enabling dependable transaction authentication, ECDSA strengthens the overall security framework of blockchains. Each transaction is validated through a distinct ECDSA signature which verifies the originator’s identity and guarantees the integrity of the transaction data. Nevertheless, current blockchain architectures that utilize ECDSA frequently encounter difficulties when tasked with verifying large volumes of signatures efficiently [11–14].

Blockchain and ZSS signatures are frequently combined to improve data integrity and authentication within distributed systems especially in cloud storage and Internet of Things (IoT) applications. ZSS signatures known for their brevity and computational efficiency are well-suited for blockchain environments that manage large volumes of data. However, a significant limitation of the ZSS (Zero-Knowledge Signature) scheme in blockchain applications is its dependence on a trusted third party (TTP) for some functions which may raise concerns regarding privacy and system efficiency. Although ZSS signatures provide benefits such as public verifiability they can encounter issues like privacy vulnerabilities and inefficiencies in particular scenarios notably when used alongside blockchain for auditing or data storage [15–19].

The growing need for secure and efficient encryption methods coupled with the necessity to overcome the limitations discussed earlier led us to develop a multi-step hybrid approach. This approach integrates TPA, an improved Hyperledger Fabric blockchain, an improved Enc-Block and upgraded versions of ECDSA and ZSS algorithms within the IoT-Cloud environment. The layered design of this framework ensures that data is processed through multiple stages each introducing a distinct enhancement. This layered integration notably bolsters the algorithm's robustness against cryptographic attacks. Furthermore, the proposed method offers benefits such as lowered time and communication overheads along with reduced computational effort for various processes.

The key innovations of the proposed multi-step (multi-layer) approach are outlined as follows:

- The first layer incorporates three cryptographic algorithms—Hyperledger Fabric, Enc-Block and a hybrid ECDSA-ZSS algorithm—that have each been individually refined. Collectively referred to as H.E.EZ, this suite of improved algorithms handles data encryption within the proposed framework. Its implementation significantly boosts security, scalability, speed and overall efficiency while also lowering both computational costs and processing time thus making it well-suited for deployment in IoT-Cloud environments.

- The second layer, known as the Credential Management Layer, is presented here as an independent component within the multi-step framework. This layer is responsible for validating data encrypted by the improved Hyperledger Fabric which is isolated from other encryption processes. Its core function is to ensure proper credential management and maintain data integrity. By doing so, the Credential Management Layer strengthens security measures by preventing fraudulent activities and safeguarding sensitive information related to users and organizations.

- The third layer, termed C-AUDIT, is dedicated to time management and auditing processes. It comprises two core components: Temporal Ordering which organizes events in a time-sensitive sequence and M-Audit responsible for managing audit channels. This layer plays a key role in minimizing system traffic, reducing time and communication overhead and lowering the computational cost of operations. Additionally, it ensures synchronized and efficient interaction between different layers of the proposed multi-layered architecture.

The rest of the article is structured in the following manner:

Section 2 provides an overview of the related works in the field. In Section 3 the proposed multi-step framework is introduced and its structural design is discussed in detail. Subsection 3.1 describes the operational workflow and the sequential steps of the framework. Subsection 3.2 defines the architectural zones of the proposed model. Subsection 3.3 elaborates on the H.E.EZ encryption layer and its functional stages. Subsection 3.4 is devoted to the definition and responsibilities of the Credential Management Layer while Subsection 3.5 outlines the structure and role of the C-AUDIT layer. Section 4 includes the performance analysis and security evaluation and Section 5 concludes the study with a summary of key findings.

## 2.Related Works

This section presents a review of prominent recent cryptographic studies that focus on one or a combination of TPA, blockchain, ECDSA or ZSS while also addressing their limitations. By conducting this analysis, our goal is to achieve a deeper insight into the advantages and limitations embedded within each approach, thereby laying the groundwork for crafting more efficient responses to present challenges.

In 2017 Liu and colleagues [20] introduced a blockchain-based framework designed to ensure data integrity in IoT environments. Their approach aimed to enable trustworthy verification of data integrity for both data owners and users eliminating the need for reliance on third-party auditors (TPAs). Despite its innovative design the framework was applicable only in small-scale use cases limiting its broader adoption.

In 2019, Xue et al. [21] proposed a novel auditing framework grounded in identity verification, specifically designed for cloud storage systems. Their approach utilized nonces recorded on a blockchain to construct challenge messages that were both unpredictable and easily verifiable. This mechanism effectively safeguarded users against tampering attempts by dishonest third-party auditors (TPAs). A comprehensive security evaluation confirmed the robustness of the scheme in preserving data integrity under various attack models. Nevertheless a notable shortcoming of the proposed method was its lack of assurance regarding the timely execution of audit processes.

In 2019 Zhang et al. [22] introduced a certificate-less public verification scheme designed to address issues caused by procrastinating auditors (CPVPA) by leveraging blockchain technology. Their approach required auditors to log each verification result as a transaction on the blockchain. Given the time-sensitive nature of blockchain transactions every verification record was timestamped at the moment of recording enabling users to verify whether auditors completed their tasks within the specified timeframe. This method effectively resolved challenges associated with certificate management. The authors supported their proposal with thorough security proofs and comprehensive performance evaluations to confirm CPVPA’s robustness and efficiency. Nonetheless the scheme faced limitations notably the absence of dynamic data update handling and relatively high computational overhead.

In 2019, Zhu et al. [23] introduced a method for ensuring data integrity, utilizing the short-signature (ZSS) technique to mitigate processing load and enhance the performance of digital signatures in RSA and BLS. The scheme supports privacy preservation and enables public auditability via a designated trusted entity (TPA). Through reducing the hash function overhead during the signing process the computational load was effectively decreased. Under the Computational Diffie-Hellman (CDH) hardness assumption the proposed scheme resists adaptive chosen-message attacks. Analyses demonstrated that this scheme outperforms RSA and BLS in terms of efficiency and security;Nevertheless, it cannot guarantee data integrity in environments with multiple replicas, nor does it accommodate multiple users.

Panjwani [24] introduced in 2017 a flexible, extensible framework for implementing ECDSA on prime fields. The study offers comprehensive guidance for deploying ECDSA using prime field sizes recommended by NIST, ranging between 192 and 521 bits. The design utilizes a combined hardware–software model on a configurable FPGA platform (Xilinx xc6vlx240T-1ff1156). Core tasks including private key creation, binary weight computation, and SHA message setup are executed in software using the Microblaze soft-core processor. The software component on the FPGA handles parameter transfer to the hardware section where signature generation and validation are performed.
This architecture achieves significant parallelism and supports high-frequency operation. Nevertheless during ECDSA computations the general-purpose processor (GPP) remains fully occupied unable to handle other tasks and the overall approach involves considerable costs and power consumption.
In 2020 Guo et al. [25] introduced a security framework tailored for consortium blockchains integrating Hyperledger Fabric with edge computing capabilities.At the core of their framework lies a CLS2 scheme, derived from a key, featuring Controllable Lightweight Secure Certificateless Signatures, designed to improve transmission efficiency without adding computational load.Compared to conventional certificateless signatures CLS2 delivered stronger security guarantees through controllable anonymity and key derivation effectively countering public key substitution and signature forgery threats. It also enabled hierarchical privacy protection a critical feature in multi-layered IoT environments.The authors confirmed the feasibility and security of CLS2 via simulations in IoT settings, supported by formal proofs based on the Random Oracle paradigm. However, the framework exhibited drawbacks in verifying real-time data correctness and depended on the presumed computational hardness of the Elliptic Curve Discrete Logarithm Problem (ECDLP) for its fundamental security

Huang et al. [26] in 2014 explored the challenges that arise when Third-Party Auditors (TPAs) in cloud environments are not fully trustworthy or may even act maliciously under specific circumstances. To tackle this issue, they proposed an innovative feedback-driven auditing framework allowing users to independently confirm the correctness of their remotely stored information without requiring direct interaction with the Cloud Service Provider (CSP) or complete dependence on the TPA.Their approach aimed to maintain data assurance while minimizing trust assumptions. Security and performance evaluations showed that the proposed method was more lightweight and secure than previous solutions. Nevertheless the model exhibited weaknesses against certain attack vectors and fell short in providing anonymity guarantees for users.
In 2016 Li et al. [27] introduced a heterogeneous signcryption-based mechanism to regulate user access behavior in IoT environments. The proposed scheme was formally verified under the Random Oracle Model ensuring its theoretical soundness. One of the key innovations of their work was enabling secure communication between users in a Certificateless Cryptography (CLC) domain and sensor nodes operating under Identity-Based Cryptography (IBC).Building on this framework, they introduced a protocol for managing access specifically designed for Wireless Sensor Networks (WSNs), with the goal of enhancing operational efficiency in IoT environments.Compared to existing approaches that rely on signcryption their solution achieved lower computational overhead and reduced energy consumption on sensor nodes. Nonetheless the scheme did not provide user anonymity and exhibited vulnerabilities to specific types of attacks highlighting areas for future enhancement.
In 2017, Fu and colleagues [28] proposed a privacy-preserving public verification framework designed for multi-user cloud data, leveraging a Homomorphic Verifiable Group Signature (HVGS). Their method improves upon prior techniques by mandating that at least t group administrators collaboratively reconstruct a tracing key, preventing any individual authority from misusing power and ensuring non-frameability.Moreover the scheme enables group members to monitor data changes using a structured binary tree and to restore the most recent valid data block if the current one is compromised. Comprehensive security analysis alongside practical experiments confirmed that the scheme is both secure and efficient. Nonetheless some limitations remain particularly regarding anonymity and resistance to certain types of attacks.

In 2018, Luo and collaborators [29] proposed an access control framework that aims to improve both security and operational efficiency for wireless sensor networks functioning across multiple domains in the Internet of Things (IoT). Their design permits an IoT participant within a Certificateless Cryptography (CLC) environment to interact seamlessly with a sensor node employing Identity-Based Cryptography (IBC), despite variations in system configurations. A significant feature of their framework is the inclusion of Known Session-Specific Temporary Information Security (KSSTIS), which many previous access control solutions lack. Performance evaluations demonstrated that the framework is suitable for wireless sensor networks operating in cross-domain IoT settings. However, certain limitations remain, particularly regarding user anonymity and resilience against specific attack vectors.

In 2020, Lu and colleagues [30] proposed a high-performance, distributed framework for verifying data integrity, implemented on Hyperledger Fabric (HF-Audit).Their method utilized Hyperledger Fabric as a communication framework, facilitating flexible assignment of the Third-Party Auditor (TPA) for every verification task. To improve scalability, they introduced an auditing protocol based on bilinear pairings and cryptographic commitments. Furthermore two TPA selection algorithms were devised to optimize performance under both complete and incomplete information scenarios. The authors validated the security of their scheme through formal proofs and assessed its practical performance. However, the proposed approach still had shortcomings, such as lacking mutual authentication, being exposed to specific attack vectors, and not ensuring user anonymity.

Vahi and Jassbi [31] introduced SEPAR in 2020, a lightweight mixed-mode cryptographic method employing a 16-bit data block and a 128-bit starting vector, designed specifically for Internet of Things (IoT) use cases.Their design integrates pseudorandom permutation functions with a pseudorandom generator to enhance security. Comprehensive security analyses alongside NIST statistical testing confirmed that this combination effectively strengthens resistance against standard cryptanalysis techniques including linear and differential attacks while offering faster encryption compared to legacy algorithms. Nevertheless the scheme falls short against certain other attack vectors performs slower than some of the latest ciphers and lacks anonymity features.In 2022, Ge and co-researchers [32] formally outlined a framework for Revocable Attribute-Centric Encryption ensuring Data Integrity (RABE-DI). Based on this groundwork, they devised a practical implementation of the RABE-DI method and thoroughly proved that it upholds both secrecy and data validity within the introduced framework.Their implementation along with an evaluation of performance demonstrated that the method is both feasible and high-performing.Nonetheless it falls short in providing anonymity, robust authentication and resilience against specific categories of cryptographic attacks.

In 2022, Alshehri together with Bamasag [33] proposed an IoT-oriented access control framework grounded in attribute specifications, referred to as AAC-IoT.To tackle prevalent security concerns their design integrates Hyperledger Fabric (HLF) as a blockchain backbone. The access control model relies on Attribute-Based Access Control (ABAC) where the quantity of attributes is dynamically selected according to the sensitivity of the data in question. Corresponding access policies are then derived based on this selection. A distinctive element of their approach is the use of fuzzy logic to define the attribute count factoring in both data type and user preferences. HLF is further employed to handle metadata and security credentials from data owners and users alike using a lightweight hashing function to ensure secure data handling. The scheme was implemented in Java and simulated using the iFogSim platform. Evaluations based on latency, throughput and storage overhead confirmed its superior performance compared to earlier methods. Nonetheless the design does not support anonymity which remains a limitation.

In 2022, Bian and colleagues [34] presented a method based on identity principles for verifying data ownership remotely, enabling the data holder to assign a specific verifier.The scheme incorporates a random integer to blind the data integrity proof enhancing privacy protection. Additionally it uses a Merkle hash tree structure to support dynamic data updates efficiently. One of the key advantages of this design is its avoidance of the complex certificate management typically required in public key infrastructures. The security guarantees rely on well-established assumptions namely the Discrete Logarithm and Computational Diffie-Hellman problems. Both theoretical analyses and experimental evaluations confirm that the scheme Is practical and effective. However it does have some drawbacks including a lack of anonymity and susceptibility to certain attack vectors.

Perera et al. [35] in 2022 introduced two widely recognized signing mechanisms, namely group-based and ring-based methods, which ensure user anonymity by hiding the user's identity within a set. Group-based methods grant conditional anonymity within a predetermined set and can be revoked, whereas ring-based methods provide persistent anonymity via spontaneously formed sets. Nonetheless, these signing mechanisms are inefficient and vulnerable to quantum-level threats and side-channel exploits. The authors emphasized that it is possible to maintain privacy while enabling traceability in group-based methods and controlled anonymity in ring-based methods, and suggested that future work should target efficiency improvements and other associated concerns.

Table 1 provides a concise overview of the methods covered in this section highlighting their respective advantages and limitations.

Table 1. Overview of Methods and Their Strengths and Weaknesses in Prior Studies

| Source | Approach | Strengths | Limitations |
|---|---|---|---|
| Liu and colleagues (2017) [20] | An IoT solution leveraging blockchain, excluding a TPA | • Data integrity in a fully non-centralized environment | • Limited to small-scale scenarios |
| Xue and colleagues (2019) [21] | A blockchain- and TPA-based scheme | • Overcome the malicious auditors' issue | • No guarantee to perform auditing duties on time |
| Zhang and colleagues (2019) [22] | A certificateless verification scheme using blockchain | • Prevent the malicious and delayed auditors | • No dynamic data update management<br>• High computational cost |
| Zhu and colleagues (2019) [23] | A ZSS-driven mechanism for ensuring data integrity in Cloud-IoT | • information accuracy<br>• confidentiality safeguarding<br>• robustness against adaptive targeted message attacks | • Cannot maintain information accuracy across multiple replica setups |
| Panjwani (2017) [24] | An improved technique for ECDSA based on a GPP, Microblaze, and a hardware accelerator | • executing all recommended field sizes by NIST<br>• high-frequency execution<br>• high level of parallelization<br>• high performance | • GPP cannot perform another activity when ECDSA is active<br>• high costs<br>• high energy consumption |
| Guo and colleagues (2020) [25] | A compact, manageable signature scheme without certificates | • regulated user anonymity<br>• confidentiality assurance<br>• protection against forged signature attacks | • Delayed verification of data correctness<br>• Security reliant on ECDLP hardness |
| Huang and colleagues (2014) [26] | A multi-TPA-based scheme to implement identical computational auditing, executing final verification duty by the user, proofing the process, and gathering feedback by TPA | • data integrity<br>• cloud data accessibility<br>• prevention of malicious TPA frame attack<br>• more secure than previous works<br>• lighter than previous works | • non-resistance to some attacks<br>• unable to preserve anonymity |
| Li and colleagues (2016) [27] | Heterogeneous sign encryption to control the access behavior of the users and WSN access control in IoT | • Protection within the Random Oracle framework | • Vulnerable to certain attacks<br>• Absence of user anonymity |
| Fu and colleagues (2017) [28] | Multi-level NPP public auditing | • non-frameability<br>• impossibility of single-party power abuse<br>• privacy protection<br>• security<br>• efficiency | • non-resistance against some attacks<br>• lack of anonymity |
| Luo and colleagues (2018) [29] | WSN access control in IoT | • user communication in a certificateless encryption environment, with a sensor node in an entity-based encryption environment<br>• security of temporary information specific to KSSTIS<br>• performance effectiveness | • absence of user privacy<br>• vulnerability to certain threats |
| Lu and colleagues (2020) [30] | A Hyperledger Fabric–based framework for verifying data integrity (HF-Audit) | • Efficiency<br>• decentralization<br>• security improvement<br>• scalability | • absence of mutual authentication<br>• vulnerability to certain attacks<br>• anonymity not ensured |
| Vahi and jassbi (2020) [31] | Creating a resource-efficient combined encryption system (SEPAR) | • security improvement<br>• speed improvement compared to some algorithms<br>• appropriate for IoT settings while remaining lightweight | • lower speed compared to some newer encryption algorithms<br>• vulnerable to certain attacks<br>• absence of user privacy |
| Ge and colleagues (2022) [32] | A protection-focused cryptographic framework using reversible attributes (RABE-DI) | • confidentiality<br>• information integrity<br>• operational efficiency | • lack of permission<br>• vulnerable to certain attacks<br>• absence of user anonymity |
| Alshehri & Bamasag (2022) [33] | An attribute-driven access management framework (ABAC) for IoT leveraging Hyperledger Fabric | • security improvement<br>• performance improvement<br>• access control | • absence of user anonymity |
| Bian and colleagues (2022) [34] | Identity-oriented scheme for verifying remote data ownership | • information consistency<br>• confidentiality safeguarding<br>• system protection<br>• operational effectiveness | • absence of user anonymity<br>• vulnerable to certain attacks |
| Perera and colleagues (2022) [35] | Analysis of two key signature methods (group vs. ring) | • guaranteeing user privacy | • lack of protection against certain quantum and side-channel attacks |
| | | | • no efficiency |

## 3. Proposed Scheme

This study proposes an innovative and optimized multi-phase lightweight cryptographic framework designed to enhance security speed efficiency and scalability simultaneously in data exchange within IoT-Cloud environments. Although key components such as the Third-Party Auditor (TPA) blockchain ECDSA and ZSS provide valuable functionalities each comes with inherent vulnerabilities and limitations that may impact overall system performance speed and security resilience. For example while blockchain can address trust issues associated with TPAs it still faces challenges related to privacy protection and verification efficiency. Additionally the open nature of blockchain introduces potential security risks and the growing number of transactions calls for solutions that maintain low latency. Most blockchain platforms continue to depend on ECDSA which can introduce considerable overhead. As transaction volumes grow the efficiency of ECDSA verification tends to decline. Additionally ZSS faces challenges related to data confidentiality and the risk of collusion between the Third-Party Auditor (TPA) cloud servers and IoT devices. In this work we address these vulnerabilities by carefully refining and integrating these critical components to develop a more robust and efficient security framework. Our approach involves enhancing cryptographic integration optimizing data classification and improving data transmission methods. Furthermore we have updated the sequential verification process for certificates and keys to better meet the stated objectives.

The proposed approach employs three distinct types of cryptography and auditing organized into three layers:

1. Cryptography Layer (H.E.EZ)
2. Credential Management Layer
3. Time Management and Auditing Layer (C-Audit)

Figure 1 illustrates that the complete architecture is organized into three distinct zones. These layers function within the cryptography and auditing segment.

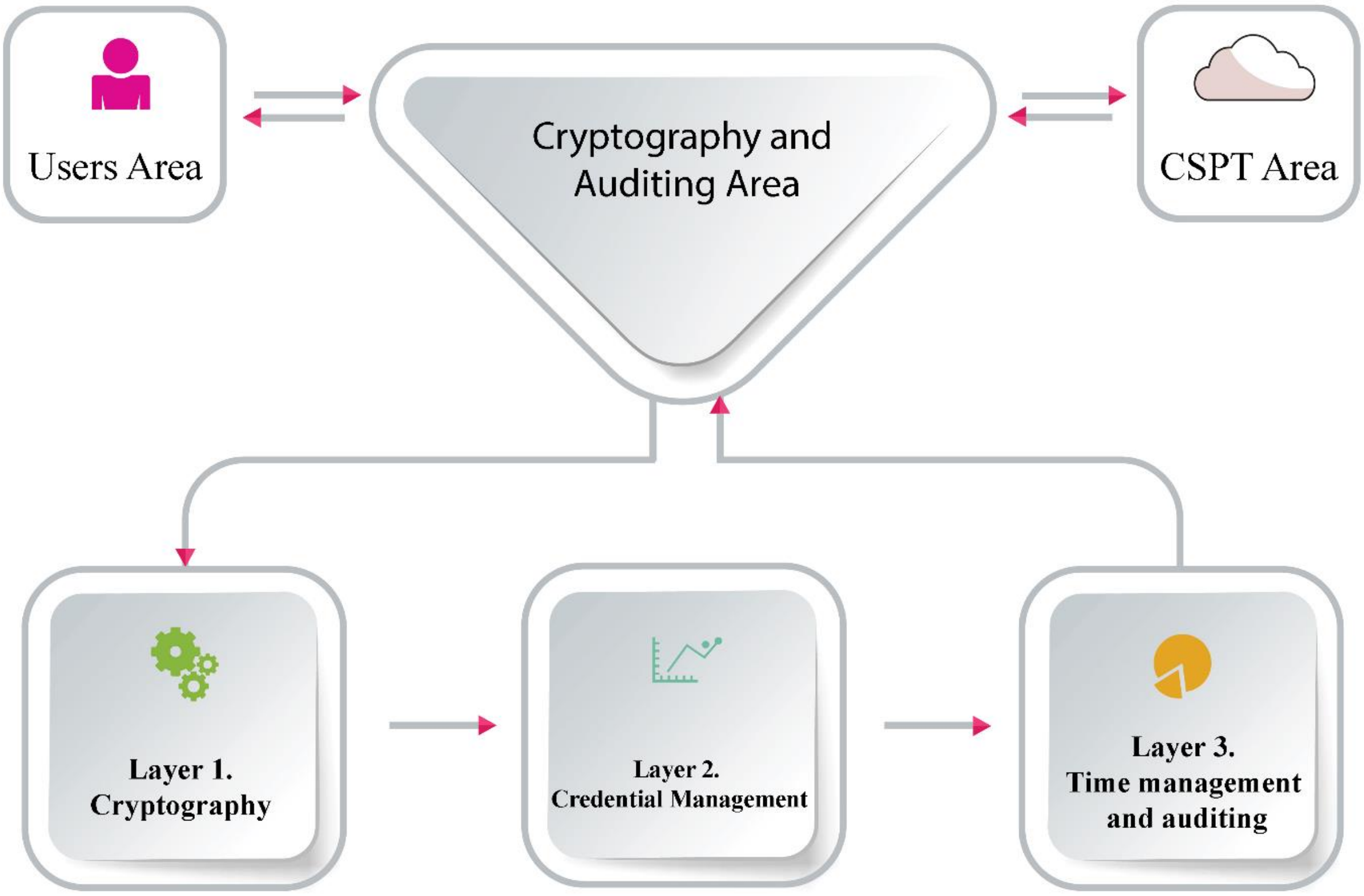

Figure 1. General overview of the proposed framework

A detailed explanation of the zones within the proposed scheme can be found in Section 3.2.

Table 2 presents an overview of the notations employed in the improved HF, EZ, and improved Enc-Block encryption schemes.

Table 2. Symbols in Improved Cryptographic Algorithms of the Proposed Scheme

| Symbols | Description |
|---|---|
| U | User |
| CSPT | Integrating Cloud Service Provider with IoT |
| $u_m$ | The participant linked to M-Audit |
| $\mathrm{CSPT}_m$ | The CSPT node linked with M-Audit |
| F | The user's document intended for upload to $\mathrm{CSPT}_m$ |
| $\mathrm{F}_I$ | The $\mathrm{i}^{\mathrm{th}}$ block of F, F=$\{F_i\}_{i\in n}$ |
| $\phi_i$ | Authentication token derived from $\mathrm{F}_\mathrm{i}$ |
| $\phi$ | A collection of authentication tokens $\phi$=$\{\phi_i\}_{i\in n}$ |
| $\mathrm{G}_1$ | A cyclic sequence set created using generator $P \in E(\mathrm{F}_q)$ with order $(q \in E(\mathrm{F}_q))$ equal to one. |
| $\mathrm{G}_2$ | A set of repeating cyclic groups having an order corresponding to $\mathrm{G}_1$ |
| $G_1$ , $G_2$ , $G_T$ | Multiple cycle groups |
| P | Primary order $G_1$ , $G_2$ |
| P | generator $G_1$ |
| G | generator $G_2$ |
| $e: G_1 \times G_2 \rightarrow G_T$ | Bilinear pairing |
| H: $\{0,1\}^* \rightarrow G_1$ | A protected hash function converting a text string into a G1 element |
| $Z_q^*$ | The collection of whole numbers ranging from 0 up to (but not including) p |
| $h: G_1 \rightarrow Z_q^*$ | A cryptographically strong hash function translating a point from G1 into an element of $Z_q^*$ |
| $PK_{CSP_m}\|SK_{CSP_m}$ | A $CSPT_m$ public-private key duo representing the user's online identity. |
| y\|x | A user-created public-private key duo for performing an audit task. |
| pk\|sk | The user produces a public-private key set for an audit task, with the secret key optionally shared at a specific phase. |
| $c_m pk\|c_m sk$ | A public-private key set created by $CSPT_m$ for auditing purposes, where the secret key $c_m sk$ may be distributed to others at a defined phase. |
| DT\|LT | A collection of TPA IDs chosen for verifying data and tag tasks |
| DTid\|Ltid | A collection of TPA IDs chosen for verifying data and tag tasks |
| VoDT\|VoLT | Verification of tasks using data and corresponding tags |
| $ic_m\|iu_m$ | Transaction ID representing the $CSPT_m$ storage, along with the user-submitted challenge |
| $sd_{bl}\|sd_{ra}$ | Audit packet for generating block count and associated random values |
| M | Specifies how many data segments need to undergo auditing |
| $E(F_q)$ | Total count of all points $(x, y) \in E(F_q)$ |
| S | Prover |
| R | Verifier |
| NI-SchnorrPoK | Utilized for demonstrating knowledge |
| ECDSA-Sign | The ECDSA signing procedures related to the IV |
| ECDSA-Verify | The validation procedures of ECDSA regarding the IV |
| Nonece | A unique value produced by CSPTm for every TPA |
| $k_i$ | 16-bit internal Enc-Block auxiliary key (sub-key)(For improved enc-Block) |
| State_i | The $i^{th}$ internal state, represented by a value from 1 to 9 (For improved enc-Block) |
| PT. | 16-bit input string for encryption (For improved enc-Block) |
| CT. | 16-bit encrypted text (For improved enc-Block) |
| $\boxplus$ | Addition operator in $2^{16}$ unit (For improved enc-Block) |
| $\boxminus$ | Subtract operator in $2^{16}$ unit (For improved enc-Block) |

## 3.1 Proposed Method Operation

The proposed method is divided into two main parts: encryption and decryption (which includes authentication and decryption). In the following sections each part will be explained in detail and further elaborated.

### 3.1.1 Encryption Phase

The encryption process begins with the user transmitting data to the encryption zone. Here three distinct algorithms are employed: an improved version of the Hyperledger Fabric blockchain, an improved Enc-Block and a refined combination of short signature and elliptic curve cryptography (EZ). Initially data is encrypted using the improved Hyperledger Fabric (Figure 2) followed by a second layer of encryption using the lightweight improved Enc-Block and EZ algorithms (Figure 3). In this dual-layer process the first few bits of the data are encrypted with the improved Enc-Block while the last few bits are secured using the EZ method. By default, the scheme assigns 16 bits to each segment, since both EZ and the enhanced Enc-Block are intended for 16-bit unit processing. This setup eliminates bitwise conversion requirements and guarantees efficient computation.

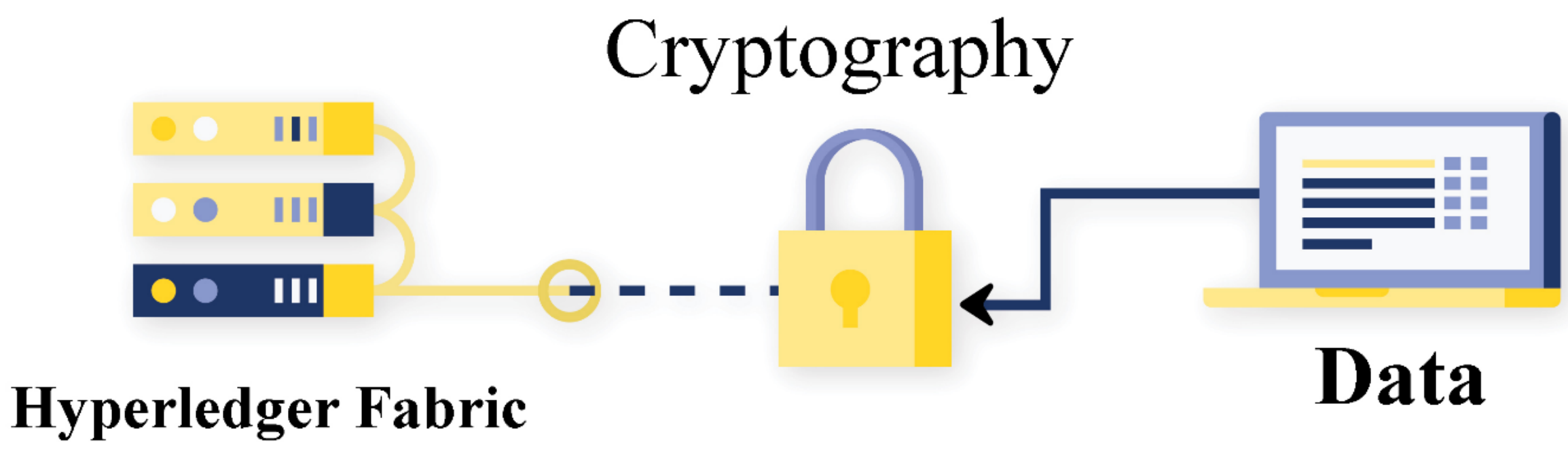

Figure 2. Encrypted data utilizing the enhanced Hyperledger Fabric platform

Cryptography layer routes the segments of data encrypted with the improved EZ and Enc-Block algorithms to the Time Management and Auditing Layer, while the remaining portion is directed to the Credential Management Layer (Figure 3).The Time Management and Auditing Layer is responsible for storing the encrypted data it receives. Meanwhile the part of the data encrypted through the improved Hyperledger Fabric that reaches the Credential Management Layer undergoes validation before being transmitted to the CSPT network.

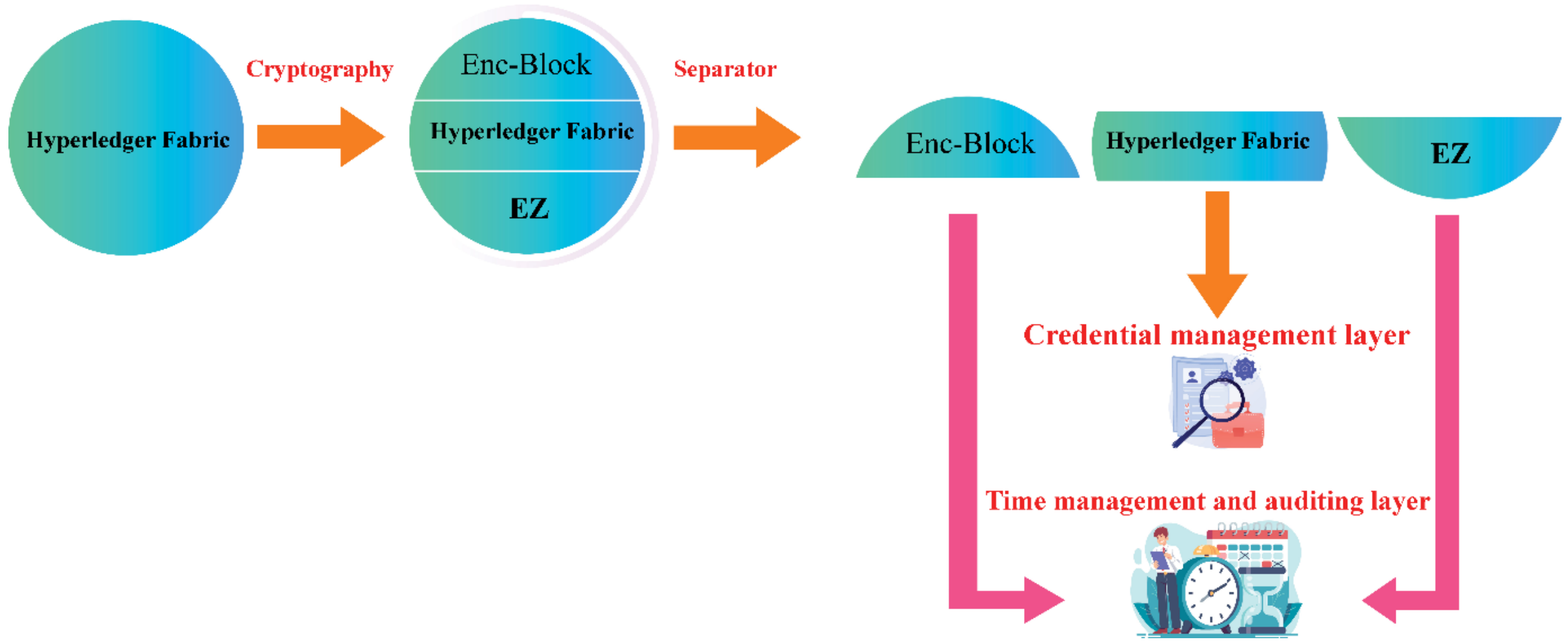

Figure 3. Steps of the encryption process in the Cryptography Layer

Pseudocode 1 presents the detailed steps of the encryption process as implemented in the proposed method.

Pseudocode 1. Overall Encryption Process in the Proposed Scheme

```
Start;
Procedure suggested method
{
Void Create area
{
Create up to 5 areas,(Users, Encryption layer, Credit management layer,C-AUDIT,CSPT,)
}
        Procedure Definition of each area
        {
                Void Users
                {
                Introduce some positions for users and refer to the user class
                 If user connected to Credit management layer then
                        Send data to HF
                        Encrypt the first 16 bits of data using enc_block encryption
                        Encrypt the last 16 bits of data using ez encryption
                                        Separate the data encrypted by separ and ez from the
        original data
                Else
                        Message "Could not connect, come back later"
                End
                }
                Void HF
                {
                Introduce some positions for HF,(crypto Hyperledger Fabric, Hyperledger Fabric
        Auditor) and refer to the HF class
                If HF connected to Users,C-AUDIT, Credit management layer then
                        Get the data from the user
                        Encrypt data using Hyperledger Fabric encryption algorithm
                        Send the encrypted data to the Credit management layer
                Else
                        Message "Could not connect, come back later"
                End
                }
                Void C-AUDIT
                        {
                Introduce some positions for C-AUDIT, Channel management, chronological
        order) and refer to the C-AUDIT class
                If C-AUDIT connected to Credit management layer then
                        Get the data from the Credit management layer
                        Send encrypted data to CSPT
                Else
                        Message "Could not connect, come back later"
                End
                }
                Void CSPT
                {
```

```
        Introduce some positions for Cloud computing network and Internet of Things and
    refer to the CSPT class
         If CSPT connected to C-AUDIT, Credit management layer then
                Get the data from the C-AUDIT
                Save the data in a safe place
        Else
                Message "Could not connect, come back later"
        End
        }
        Void M-Audit
        {
        Introduce some positions for Management of audit channels and refer to the ,C-
    Audit class
         If M-Audit connected to Credit management layer, Users,CSPT then
                Manage connections
                Manage audit channels
        Else
                Message "Could not connect, come back later"
        End
        }
                }
                        }
End;
```

### 3.1.2 Decryption Phase

Before receiving data from the CSPT network, user authentication is required. Accordingly, the first step in the decryption phase is to verify the identity associated with the data. At this stage, C-AUDIT takes precedence by establishing communication with the user. The process begins by validating the EZ algorithm, through which C-AUDIT confirms the user's identity. Subsequently, C-AUDIT authenticates the improved Enc-Block key. Once the key is validated, decryption continues for information secured through both the improved Enc-Block method and the EZ scheme. Finally, C-AUDIT merges the decrypted segments—including data encrypted via the improved Hyperledger Fabric—by concatenating the initial and final bits to reconstruct the original complete data. Once the data segments are concatenated, C-AUDIT establishes a protected link connecting the user with CSPT within the auditing channel management system, referred to as M-Audit.The verification of information secured through the improved Hyperledger Fabric framework is initiated via this channel. After a successful check, the process of deciphering starts.

Pseudocode 2 illustrates the step-by-step procedure for deciphering in the suggested approach.

Pseudocode 2. Step-by-Step Deciphering Procedure in the Suggested Approach

```
Start;
Decoding procedure of the proposed method
{
Void Connections
{
        Procedure Definition of each area
        {
            Void Users
            {
        If user connected to C-Audit layer, M-Audit, CSPT then
            Send data from CSPT to M-Audit And M-Audit checks the data
        {
            If the data is complete then
            Create a secure communication channel for the H.E.EZ layer
            Else
                Message "The data is incomplete."
        }
        Else
            Message "Could not connect, come back later"
        }
            }
            Void C-Audit layer
            {
            If M-Audit has created a communication channel related to the user
        then
                {
                Check the EZ key associated with the user
                If key is True then
                    {
                    Perform EZ decryption operation
                    Check the Enc-Block key associated with the user
                    If key is True then
                        {
                    Perform Enc-Block decryption operation
                    Link EZ and Enc-Bolck data with data received in M-
                Audit
                    M-Audit creates another communication channel for
                the user and CSPT
                    M-Audit sends the completed data to the H.E.EZ
                    layer through the communication channel it has
                    previously created
```

```
                    Else
                            Message "The key is wrong."
                            }
                                    }
            Else
                    Message "Channel does not exist."
            }
                    }
            Void H.E.EZ layer
            {
            Get the data from the channel that M-Audit created.
            If the received data is complete thn
                    {
                            Contact the administrator and have them perform the
                            Hyperledger Fabric decryption operation and deliver the
                            decrypted data to the relevant user.
            Else
                    Message "The data is incomplete."
                    }
                            }
                                    }}

End;
```

## 3.2 Zones of the Presented Framework

To examine the zones within the presented framework, we refer back to Figure 1, which highlights three separate areas. These zones are additionally depicted in the block diagram shown in Figure 4.

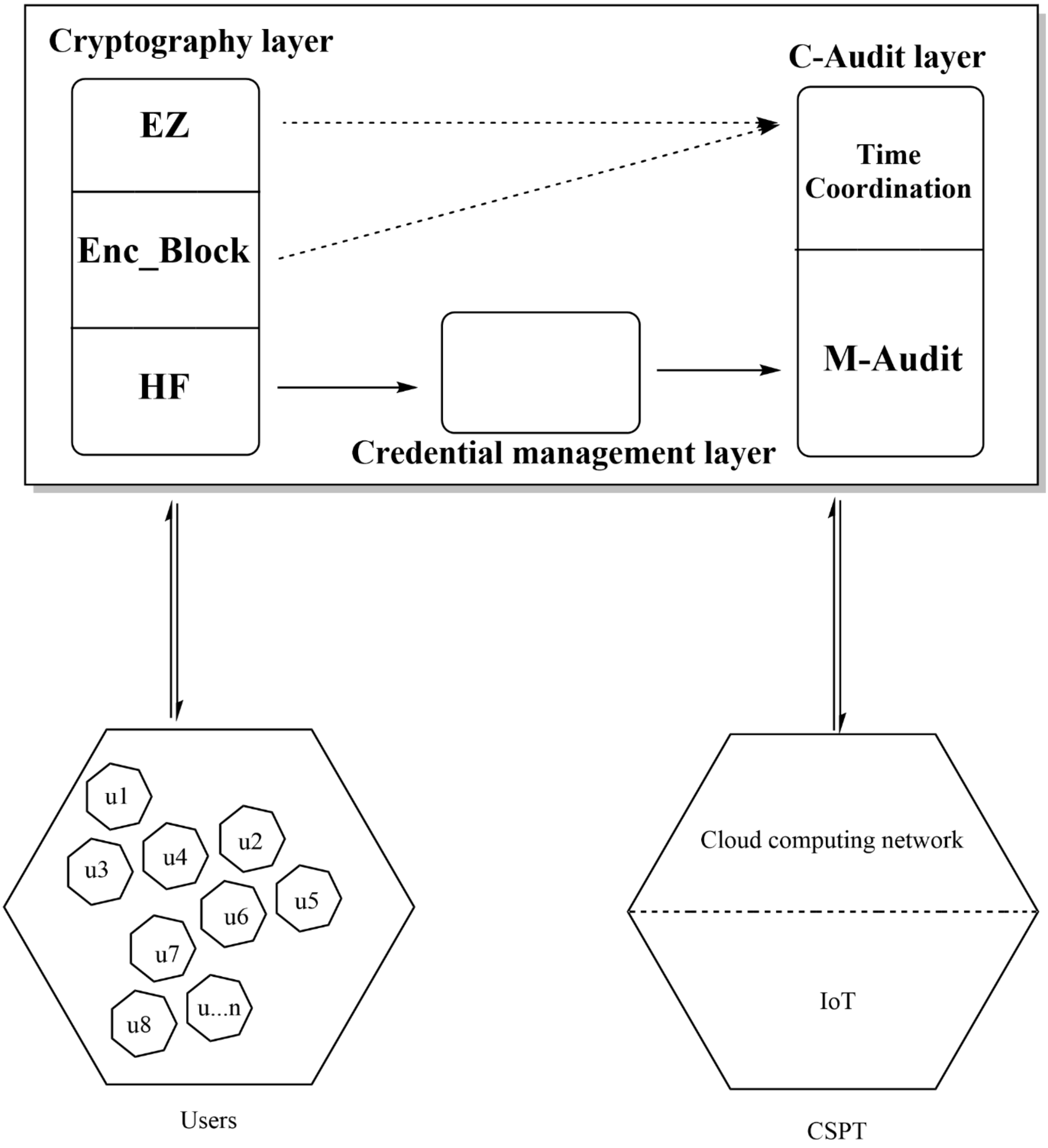

Figure 4. General block diagram of the proposed approach

The zones mentioned above are detailed in the subsequent sections.

### 3.2.1 Zone 1: Users

Zone 1 encompasses users, which can be either individuals or organizations that delegate their data to cloud platforms and IoT systems to meet storage and computational demands while minimizing costs. Within this framework, users transmit their data to the Cryptography and Auditing Zone for encryption and communicate with the Time Management and Auditing (C-AUDIT) layer for authentication tasks.

### 3.2.2 Zone 2: Cryptography and Auditing Layer

This zone is composed of three layers:

1 .Cryptography Layer (H.E.EZ): Data is encrypted here using three algorithms — the improved Hyperledger Fabric, the improved Enc-Block and the refined EZ scheme

2. Credential Management Layer: Responsible for authenticating data encrypted by the improved Hyperledger Fabric, which is isolated from the other encryption algorithms.

3. Time Management and Auditing Layer (C-AUDIT): Includes components for time coordination and the management of auditing channels (M-Audit).

### 3.2.3 Zone 3: Network for Cloud Platforms and IoT Systems

As suggested by its title, this zone embodies a unified connection between cloud platforms and IoT systems, providing users with various services. These services are generally customized based on user requirements and requests, with associated costs varying accordingly. For simplicity, we refer to this combined network as CSPT.

### 3.3 Overview of the H.E.EZ Cryptography Layer (First Layer) in the Proposed Scheme and Its Workflow

In this section, we present the improved cryptographic methods applied in the proposed approach and detail the operational procedures of each of the three encryption algorithms.

#### 3.3.1. Improved Cryptographic Framework of Hyperledger Fabric (HF)

In the hybrid model presented here, the Hyperledger Fabric (HF) architecture is conceptually inspired by [30] while integrating substantial enhancements that distinguish it from the original framework. As described in [30], the HF-Audit system includes users, a Cloud Service Provider (CSP), and a Third-Party Auditor (TPA), each responsible for specific roles within the system.Fabric Certificate Authority (Fabric-CA), an administrator, and two separate channels for auditing and credential management—designed primarily for cloud-based environments. In contrast, our architecture decouples several of these components—namely, the users, audit channel, credential channel, and TPA—from the HF-Audit module, redefining them as independent entities. This redesign enhances the system's modularity, scalability, and adaptability while reducing overall computation overhead and improving processing speed. As a result, energy consumption is also optimized. Furthermore, unlike the model in [30], the improved HF framework introduced here is fully compatible with the operational demands of IoT–Cloud ecosystems. Figure 5 illustrates the internal structure of the proposed improved HF system.

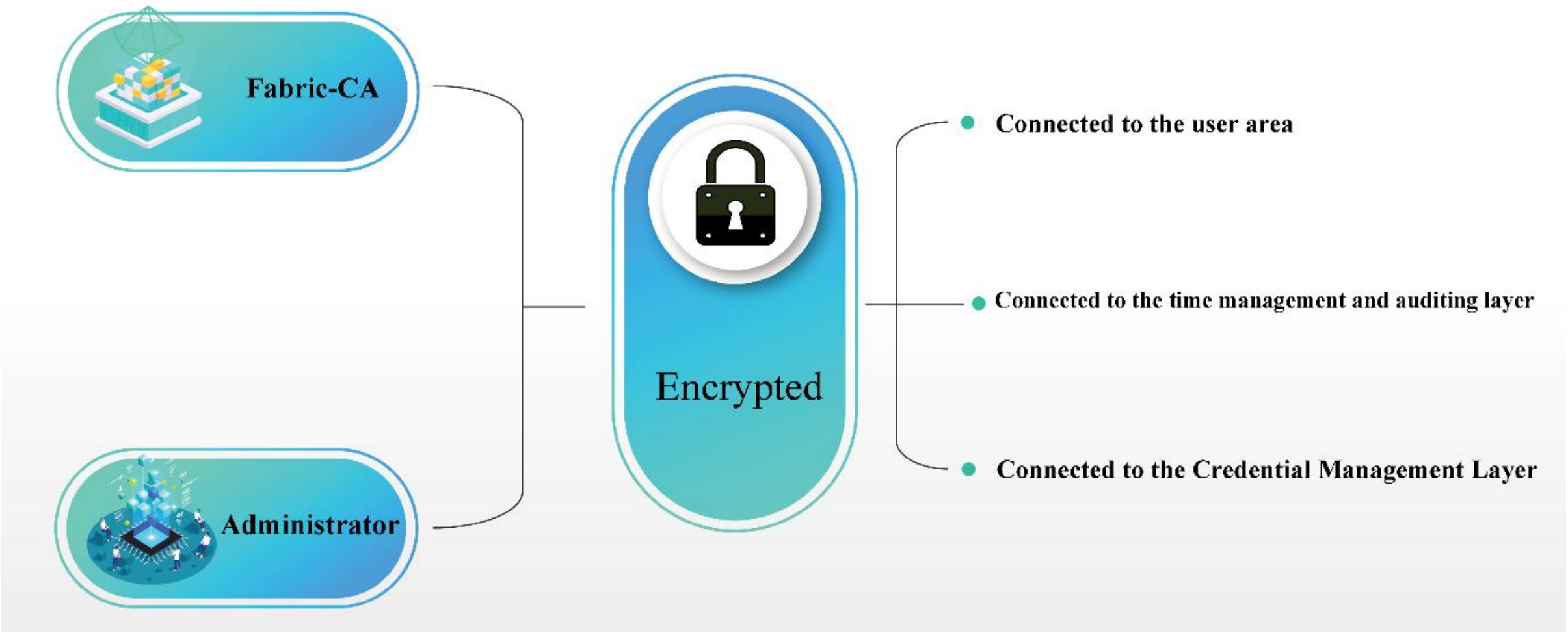

Figure 5. Step-by-Step Encryption Process in the Improved Hyperledger Fabric (HF)

Each phase's role in the improved Hyperledger Fabric (HF) within the proposed framework is outlined as follows:

**1.Registration:** In this step, every participant must enroll in the consortium via Fabric-CA. Once the administrator verifies and approves the registration, the participant gains permission to join the channel. To maintain transparency and enable effective communication each participant's identity is shared with other members of the network.

**2.Setup**: At this stage, $CSPT_m$ (Table 2) publishes the required cryptographic parameters.Using these inputs, the user creates a single-use public and private key pair for auditing. The parameters include the cyclic groups $G_1$, $G_2$, and $G_T$. The symbol p indicates the prime cardinality of $G_1$ and $G_2$, while g serves as a base element for $G_2$, The mapping $e: G_1 \times G_2 \rightarrow G_T$ specifies the bilinear map. The function $H: \{0,1\}^* \rightarrow G_1$ transforms binary strings of any length into elements of $G_1$,and $h: G_1 \rightarrow Z_q^*$ is another mapping that converts points in $G_1$ into members of to elements in $Z_q^*$.

Finally, the user selects a random secret key $X \in Z_q^*$ and computes the corresponding public key $y = g^x \in G_2$(Table 2).

**3.Storage phase:** At this stage, the user prepares the necessary inputs and transmits them off-chain to $CSPT_m$.Subsequently, $CSPT_m$ aggregates these inputs into a transaction record and records it on the ledger.

- File Delivery: Initially, the user splits F into n blocks ($F = F_i( i \in n)$. Afterwards, a random value $u_m \in G_1$ is produced. For each block, the value

  $\phi_i = (H(F_i) \cdot u_m^{(Fi)})^x$ is calculated, forming the set $\phi = (\phi_i \mid i \in n)$.

  The user transmits the following data to $CSPT_m$ (refer to Table 2):

  $\left\{\phi,F,n,g,y,e,h,\{ID,u_m\}_{PK_{CSPT_m}},\{g||u_m\}_{SK_{User}}\right\}$.

  This process ensures secure and verifiable delivery of the file blocks to the $CSPT_m$ entity.

- Storage: Upon receiving the message, $CSPT_m$ retrieves the user's identifier to confirm their registration status. If the user is not registered, the process ends. Otherwise, $CSPT_m$ validates the encrypted content by computing $H(F_i)$ $(i \in n)$ for every block $F_i$. Next, a random key pair ($c_m pk$, $c_m sk$) is generated, and $\{c_m pk\}_{PK_{user}}$ is produced and then utilized. Following this, a transaction containing

  $\mathit{The\ set}\ \{H(F_i)(i\in n),n,g,y,e,H,h,u_m,\{c_m pk\}_{PK_{user}}\}$
  is recorded on the ledger.Subsequently, a local entry $SR =$ $(ID, F, c_m pk, c_m sk, ic_m)$ is created, where $ic_m$ denotes the transaction identifier allocated by Fabric (see Table 2).

### 3.3.2. Improved Lightweight Encryption Using the Enc-Block

Once the data has been encrypted via the improved Hyperledger Fabric (HF), the process proceeds with applying the refined Enc-Block encryption. In this stage, a specific segment of the encrypted data—namely, the initial 16 bits produced by the HF encryption—is subjected to an additional layer of encryption using the improved Enc-Block method. This section outlines the implementation and role of the Enc-Block encryption within the proposed architecture.

Drawing on the SEPAR encryption framework [31], we have isolated and improved the Enc-Block component to better suit the requirements of our system. While SEPAR is known for its efficiency and presents several notable strengths, it demonstrates slower performance compared to some other encryption schemes discussed in prior literature. To overcome this limitation, our proposed multi-layer encryption architecture incorporates SEPAR selectively, employing it within a broader, more adaptive encryption model aimed at improving speed and overall performance. This study seeks to retain the core advantages of SEPAR while addressing its weaknesses—most notably, its slower processing time. Accordingly, only the efficient features of SEPAR's Enc-Block are utilized, whereas the remaining elements are intentionally left out. This selective integration allows our design to achieve reduced computational load, leading to faster, more lightweight encryption without compromising security. The improved Enc-Block encryption process employs eight 16-bit blocks, referred to as Enc-Blocks. Each block incorporates an internal b16 block cipher function. The process begins with an initialization algorithm, in which eight randomly generated 16-bit values are assigned to eight internal registers. This initialization algorithm then executes four successive rounds. Upon completion, the resulting eight states are moved to the internal registers of the primary encryption process. Additionally, the last output of the fourth round after setting its seventh bit to one is stored in the ninth register of the encryption process [31].

Pseudocode 3 illustrates the operational process of the improved Enc-Block algorithm as applied in the proposed framework.

Pseudocode 3. Initialization and encryption process of the improved Enc-Block

```
Start;
Input Data
Procedure Enc-Block
{
Void Data wrangling
{
Input: eight 16-bit random numbers (Nonce)
Output: eight states with initial value and a linear transition constant
state_1 = NONCE_1
state_2 = NONCE_2
```

$state_3$ = $NONCE_3$

$state_4$ = $NONCE_4$

$state_5$ = $NONCE_5$

$state_6$ = $NONCE_6$

$state_7$ = $NONCE_7$

$state_8$ = $NONCE_8$

**for** t = 0 to 3 **do**

$V12_t = Enc_Block_{k1}(((state1_t \boxplus state3_t) \boxplus state5_t) \boxplus state7_t)$

$V23_t = Enc_Block_{k2}(V12_t \boxplus state2_t)$
$V34_t = Enc_Block_{k3}(V23_t \boxplus state3_t)$
$V45_t = Enc_Block_{k4}(V34_t \boxplus state4_t)$
$V56_t = Enc_Block_{k5}(V45_t \boxplus state5_t)$
$V67_t = Enc_Block_{k6}(V56_t \boxplus state6_t)$

$V78_t = Enc_Block_{k7}(V67_t \boxplus state7_t)$

$Out_t = Enc_Block_{k8}(V78_t \boxplus state8_t)$

$state1_{t+1} = state1_t \boxplus Out_t$

$state2_{t+1} = state2_t \boxplus V12_t$

$state3_{t+1} = state3_t \boxplus V23_t$

$state4_{t+1} = state4_t \boxplus V34_t$

$state5_{t+1} = state5_t \boxplus V45_t$

$state6_{t+1} = state6_t \boxplus V56_t$

$state7_{t+1} = state7_t \boxplus V67_t$
$state8_{t+1} = state8_t \boxplus V78_t$
**end for**

LFSR = $Out_3$ | 01000

**return** $statei_7$ ($i$ = 1. …. . 8) and LFSR

}
Void Encrypt data
{
If the data is wrangled correctly then
Go to next step
Else
Return step 7
End if

Input: 16-bit plain text and eight 16-bit modes
Output: 16-bit cipher text

$V12_t = Enc_Block_{k1}(PT_i \boxplus state1_t)$

$V23_t = Enc_Block_{k2}(V12_t \boxplus state2_t)$

$V34_t = Enc_Block_{k3}(V23_t \boxplus state3_t)$

$V45_t = Enc_Block_{k4}(V34_t \boxplus state4_t)$

$V56_t = Enc_Block_{k5}(V45_t \boxplus state5_t)$

$V67_t = Enc_Block_{k6}(V56_t \boxplus state6_t)$

$V78_t = Enc_Block_{k7}(V67_t \boxplus state7_t)$

$CT_i = Enc_Block_{k8}(V78_t \boxplus state8_t)$

$LFSR_{t+1} \leftarrow LFSR_t$

$state2_{t+1} = V12_t \boxplus V56_t \boxplus state6_t$

$state3_{t+1} = V23_t \boxplus state4_{t+1} \boxplus state1_t$

$state4_{t+1} = V12_t \boxplus V45_t \boxplus state8_t$

$state5_{t+1} = V23_t \boxplus LFSR_{t+1}$

$state6_{t+1} = V12_t \boxplus V45_t \boxplus state7_t$

$state7_{t+1} = V23_t \boxplus V67_t$

$state8_{t+1} = V45_t$

$state1_{t+1} = V34_t \boxplus V23_t \boxplus V78_t \boxplus state5_t$

**return** $CT_i$

}
}
End;

### 3.3.3. EZ Encryption (Improved Scheme)

Following the encryption of the initial bits via the improved Enc-Block method, the last 16 bits undergo encryption via a hybrid encryptor called EZ—an improved combination of ECDSA and ZSS [36].This upgraded EZ scheme relies on bilinear pairings, short signatures, commitments and cryptographic protocols grounded in zero-knowledge proofs, all constructed upon elliptic curve cryptography. Elliptic curves, as noted in [37], provide highly efficient bilinear pairings that are well-suited for verifying the integrity of committed data within blockchain environments. The procedure for bilinear pairing computations has been detailed earlier.

The hybrid system architecture described in [36], which integrates elliptic curves and ZSS within a blockchain framework, involves four key entities: the user, identity validator, certificate provider and service provider. To improve scalability and expand the design, our proposed approach excludes the user and identity validator entities from the EZ domain. Instead, a newly introduced Credential management layer, along with M-Audit in the C-Audit layer, serves as a bridge linking the EZ domain with the user's domain, now relocated beyond the EZ perimeter. Figure 6 illustrates the schematic layout of the EZ's architectural design inside the suggested multi-layered framework.

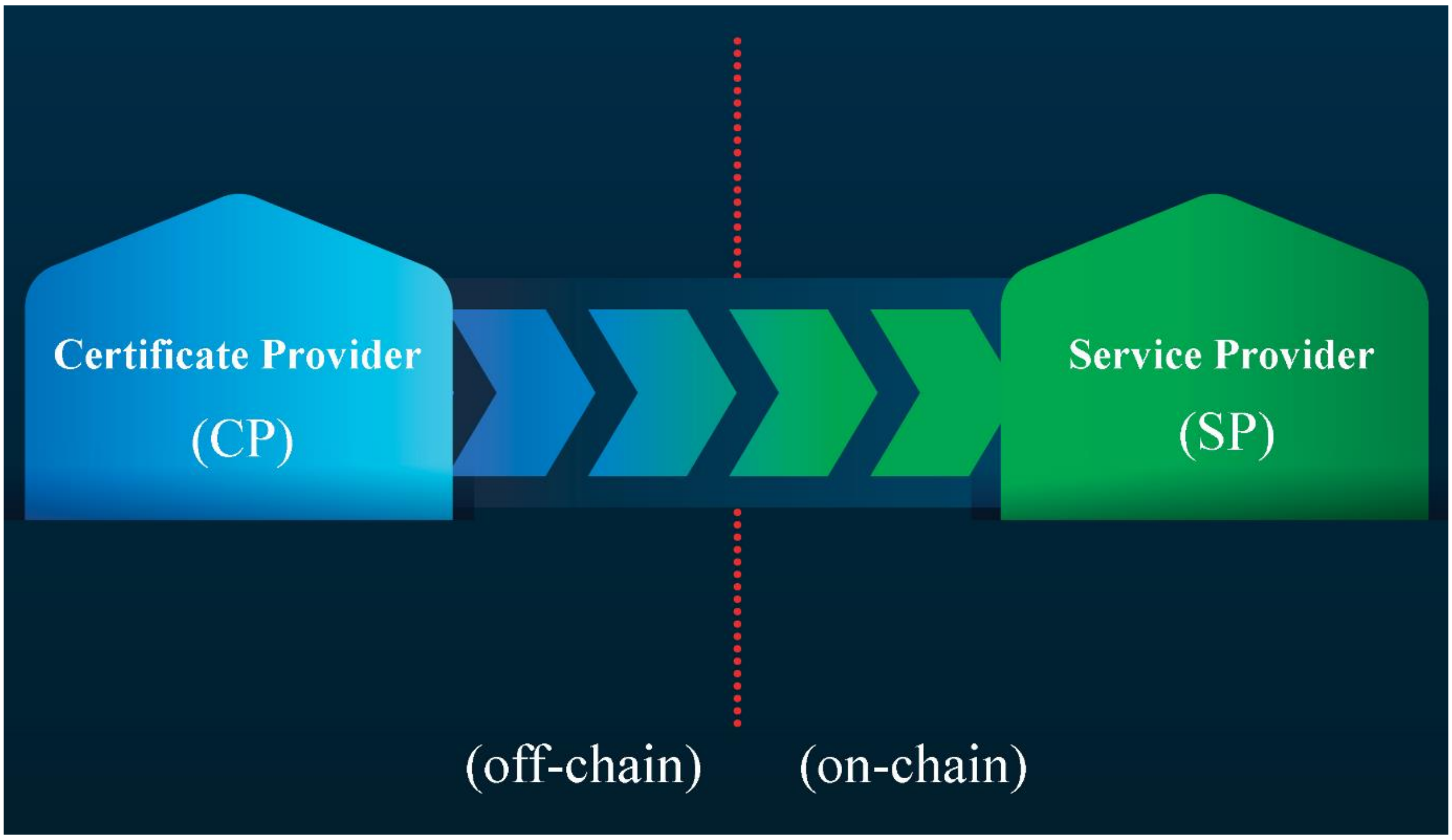

Figure 6. EZ System Architecture within the Suggested Framework

The roles of components in the EZ architecture as proposed (Figure 6) are outlined below:

- **Certificate Provider (CP):** Upon anonymously verifying the identity and confirming the attributes of user $u_m$, the CP grants a tailored credential for each service. This credential plays a key role in managing both security and privacy. Equipped with it, user $u_m$ gains access to blockchain services and can carry out activities such as online transactions.

- **Service Provider (SP):** The SP delivers specific blockchain services to each verified and authorized user $u_m$. Upon request, the SP checks and validates the credentials presented by user $u_m$ to access chain services. Importantly, this validation is performed without examining the user's identity or attribute details. Generally, the SP requires user $u_m$ to adhere to certain predefined restrictions.

In the proposed framework, EZ facilitates both on-chain and off-chain communications (Figure 6). Off-chain communication involves interactions conducted without recording data on the blockchain ledger. The EZ protocol suite within this design is composed of the three following components:

- **Commitment Mechanism**
  The described protocol operates as an off-chain mechanism and is primarily intended for generating cryptographic signatures and confirming the identity-related attributes of user $u_m$ through general claim verification. In this process, the Identity Validator (IV), which is integrated into the Credential management layer of the proposed system, cross-examines the user's declared attributes, submitted claims, and supporting documentation. It evaluates a defined attribute set — $v_1$, $v_2$, …, $v_m$— to verify the corresponding values X and determine the legitimacy of $u_m$ as an eligible participant. Once validation is complete, $u_m$ utilizes cryptographic procedures to: (1) construct a commitment over the verified data, and (2) obtain an ECDSA-based signature from the IV, thereby finalizing and securing the commitment C (Figure 7).
- Key Generation Setup for $u_m$ in the Commitment Protocol:
  Let G be a cyclic group with prime order q and generators $P_0$, $P_1$, $P_2$, ..., $P_n$ ϵ $E(F_{||})$. Two keys are generated: a secret key denoted as $sk_{u_m}$ and a public key defined by $pk_{u_m} = sk_{u_m}.P_0$. The resulting output of this setup is the tuple $\left(E(F_{||}), q, (P_0, P_1, P_2, \dots, P_n), pk_{u_m}\right)$. To implement the IV signature, the ECDSA algorithm is utilized, a widely recognized encryption standard in blockchain systems.The key generation, signing, and verification procedures for ECDSA related to IV are respectively referred to as ECDSA-KeyGen, ECDSA-Sign, and ECDSA-Verify.As illustrated in Figure 7, $u_m$ obtains the ECDSA-Sign over the committed attribute values C from IV, facilitating anonymous authentication on the blockchain.

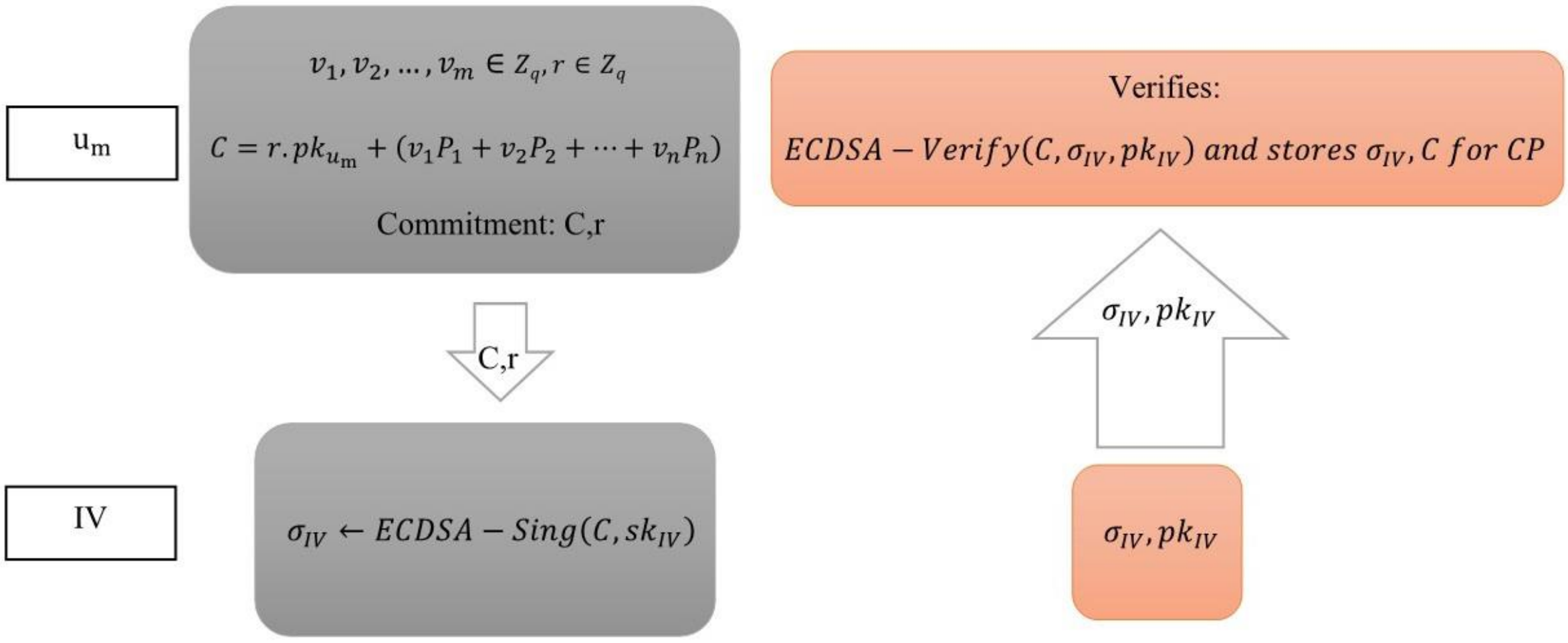

Figure 7. Signature issuance in the EZ commitment mechanism of the suggested scheme

- **Credential issuance procedure**

  To access services on the blockchain, $u_m$ communicates with the Certificate Provider (CP) via the Credential management layer to confirm eligibility for credential issuance. This process can take place off-chain, though on-chain implementation is also possible. The CP reviews the identity attributes of each $u_m$ to assess their qualification. Since this verification might conflict with privacy regulations, $u_m$ provides a valid signed commitment from the Identity Verifier (IV) accompanied by a non-interactive Schnorr knowledge proof (NI-Schnorr PoK, Table 2) that conveys all necessary data to reconstruct the commitment securely and prevent misuse within the network. The CP then authenticates both the IV's signature and the accompanying proof, and may further validate particular claims. Once verified, the credential is granted to $u_m$, enabling efficient and secure interaction with various Service Providers (SPs) across the blockchain.

The credential issuance procedure within EZ, as outlined in the proposed scheme, is detailed in Pseudocode 4.

Pseudocode 4. Procedure for Credential Issuance in EZ under the Proposed Framework

Start;<br>
Procedure Credit issuance protocol<br>
{<br>
Void u<br>
{<br>
Input: $C = r.pk_u + (v_1P_1 + v_2P_2 + \cdots + v_nP_n), (v_1, v_2, .., v_n) \in Z_q, \sigma_{IV}$<br>
//Calculate EC based NI-Schnorr PoK of C<br>
Select: $w \in_R Z_q^*$<br>
Calculate: $A = w.pk_u, s, H(A)\ and\ t = s.r + w$<br>
 For $i \in [1, n]$<br>
  Choose: $w_i \in_R Z_q^*$<br>
  Calculate: $A_i = w_i.P_i, s_i, H(A_i)\ and\ t_i = s_i.r_i + w_i$<br>
End for<br>
Send: $C, \sigma_{IV}, \left(A, t, r.pk_{u_m}\right), (A_i, t_i, v_i, P_i)$ to CP<br>
}<br>
Void CP<br>
{<br>
Verifies: $ECDSA - Verify(C, \sigma_{IV}, pk_{IV})$<br>
If verified then<br>
 Verify $t.pk_{u_m} = A + r.pk_{u_m}.s, t, P_i = A_i + v_i.P_i.s_i\ and\ C = r.pk_{u_m} + v_i.P_i$<br>
 {<br>
  If verified then<br>
   CP signs $M_{ZSS-Sign}(C, sk_{CP})$: $computes\ \sigma_{CP}(H(C) + sk_{CP})^{-1}pk_{u_m}$<br>
  Else<br>
   Message ”u not verfity”<br>
  End<br>
 }<br>
Else<br>
 message ”u not verfity”<br>
End<br>
Send: $C, \sigma_{CP}, pk_{CP}$ to $u_m$<br>
}<br>
Viod u<br>
Verifies: $M_{ZSS-Verify}(C, \sigma_{CP}, pk_{CP})$<br>
If verify then<br>
 $e(H(C)P + pk_{CP}, \sigma_{CP}) = e\left(P, pk_{u_m}\right)$<br>
Else<br>
 End<br>
End<br>
}<br>
 }<br>
End;

- **Credential Presentation Protocol**

This protocol differs from the previous two by being implemented directly on the blockchain. In this approach, we adapt the ZSS short signature scheme [38] to ensure the credentials are unlinkable. Utilizing elliptic curve cryptography, the design allows users to hide their credentials effectively, preventing unauthorized tracing or association. To accomplish this, we utilize the Verheul method [39] to achieve self-blinding and unlinkability of credentials. Importantly, $u_m$ is able to perform this unlinking independently, since Verheul inherently ensures these signatures cannot be linked. At the outset, $u_m$ blinds the keys $sk_{u_m}$ and $pk_{u_m} = sk_{u_m}.P_0$, together with the credential obtained from CP. Afterward, $u_m$ forwards the blinded output toward the SP and proves possession of the concealed $sk_{u_m}$ key by employing an NI-Schnorr proof of knowledge built upon elliptic curve cryptography.Despite applying blinding, SP verifies the blinded values using the appropriate verification equation, ensuring that the signature remains valid and intact.

Pseudocode 5 outlines the process of credential presentation in EZ as part of the proposed scheme.

Pseudocode 5. Credential Presentation in EZ within the Proposed Scheme

```
Start;
Procedure Credit presentation protocol
{
Void u
{
Input: σ_CP, C, sk_um, pk_um
Choose: b ∈_R Z_q^* as blinding factor compute, sk'_um = b.sk_um, pk'_um = b.sk'_um.P, σ'_CP =
b.σ_CP, P' = b.P, pk'_CP = b.pk_CP, C' = b.H(C)
//Calculate NI-Schnorr PoK of sk'_um = b.sk_um
Choose: r' ∈_R Z_q^*
Calculate: R' = r'.P, s' = H(R') and t' = s'.sk'_um + r'
Send (R', s', t') for NI-Schnorr PoK of sk'_um and σ'_CP, pk'_um , pk'_CP , P', C' for credential
validation to SP
}
Void SP
{
Choose: λ ∈_R Z_q^*
Calculate: σ = λ.pk_CP and pk''_CP = λ.pk'_CP
Send:pk''_CP to u_m
}
Void u
{
Calculate:  pk'''_CP = b^{-1}pk''_CP
```

```
Send: $pk_{CP}^{'''}$ to SP
}
Void SP
{
If $\sigma \equiv pk_{CP}^{'''}$ then
        $pk_{CP}^{'}$ is correct
        Verify $e\left((C^{'}P + pk_{CP}^{'}), \sigma_{CP}^{'}\right) = e(P^{'}, pk_{u_m}^{'})$
        $t^{'}.P = R^{'} + pk_{u_m}^{'}.s^{'}$ to proof the correctness of $sk_{u_m}^{'}$
Else
        message "u not verify"
End
}
        }
End;
```

EZ strengthens blockchain security even when processing large volumes of data, and by optimizing encryption speed, it effectively lowers the computational overhead in the proposed approach.

### 3.4. Credential Management Layer (Tier Two) within the Suggested Framework

This part presents the Credential Management Layer, functioning as the second tier within the suggested framework.It comprises several integrated components: (1) Identity Validator (IV), (2) Credential Channel, (3) Credential Ledger, (4) Third-Party Auditor (TPA), (5) Validation Operations, and (6) Audit Ledger—all built upon the cryptographic foundations of the H.E.EZ layer. This layer maintains direct interaction with M-Audit within the C-Audit layer (Figure 8). Its main objective is to create a stable and trustworthy environment for managing credentials and verifying data integrity. By effectively preventing fraudulent activities and safeguarding user and organizational information from leaks, this layer plays a critical role in strengthening overall system security.

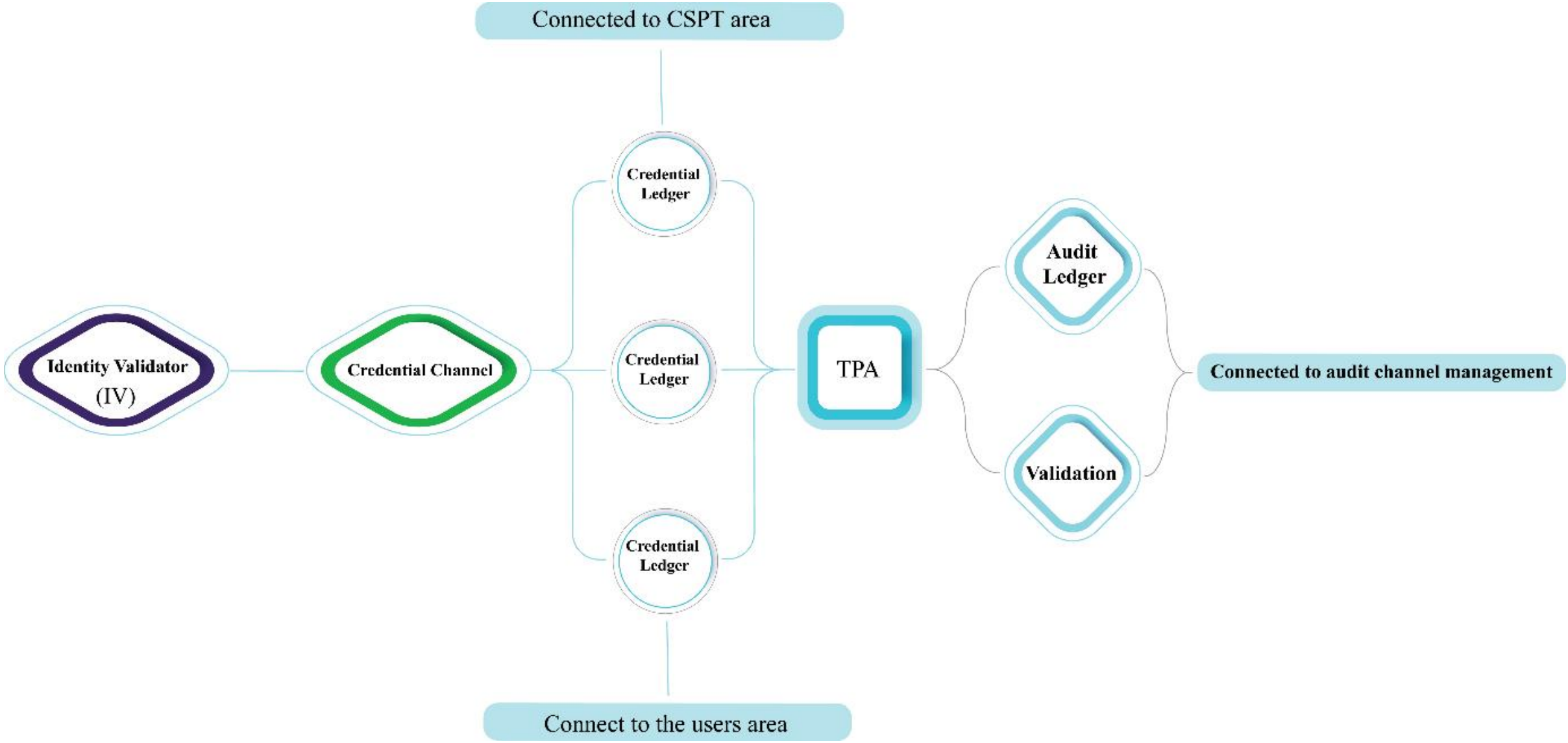

Figure 8. Credential management layer of the proposed scheme

This part describes how the Credential Management Layer interacts with M-Audit to carry out data auditing.This process leverages bilinear pairings to ensure data integrity and employs commitments based on asymmetric encryption to securely handle the identities of Third-Party Auditors (TPAs).

To begin, we first clarify the roles of bilinear pairings, commitments, and the protocol involved in the communication between the Credential Management Layer and M-Audit. These components collectively ensure that data is handled and verified with precision and security.

1.Bilinear Pairing: The bilinear map, referred to as e (Table 2), must fulfill the properties outlined in Equations (1) and (2).

(1) Bilinearity

For $u_m \in G_1, v \in G_2$ and $\forall a,b \in Z_q^*$, $e(u_m^a, v^b) = e(u_m, v) = e(u_m, v)^{ab}$

(2) Non-degenerate

$\exists g_1 \in G_1$ و $\exists g_2 \in G_2$ such that $e(g_1, g_2) \neq 1.3)$

2.Commitment. This mechanism is characterized by three fundamental properties:

**a. Integrity:** When both the sender and receiver perform their roles honestly in the interaction between the Credential Management Layer and M-Audit, the receiver obtains exactly the information that the sender committed to disclose, ensuring precise and trustworthy data transmission.

**b. Confidentiality:** Prior to the disclosure phase, the receiver has no access to any information, ensuring that data is released only securely and exactly when intended.

**c. Binding:** Once the commitment phase is complete, the sender cannot modify the promised information. This property upholds trust within the communication, ensuring that all parties adhere strictly to their commitments.

3.Channel: A fundamental element within the Credential Management Layer, each channel is composed of the following components:

- Participant (Entity)
- Anchor Node (an appointed delegate of the entity in charge of coordinating with specialized nodes)
- Record Book (kept solely for this network)
- Smart Contract Module (the program deployed exclusively for the network)
- Specialized nodes can engage in multiple networks and handle their communications across distinct record books, guaranteeing complete separation among them.In the proposed framework, two distinct channels are defined within the Credential Management Layer: one focused on auditing and the other on Credential management. The auditing channel, deployed through M-Audit, maintains records of audit-related activities including user audit requests, data storage, proofs, and verifications issued by $CSPT_m$ entities, alongside confirmations provided by TPAs. All nodes within this channel possess the capability to independently verify audit outcomes.On the other hand, the Credential channel is responsible for maintaining records concerning the credentials of TPAs. Joining this channel is mandatory for all $CSPT_m$ and TPAs, whereas for ordinary users, it remains optional. In scenarios where definitive evidence is present, the manager functions as the exclusive peer authorized to submit TPA-related information (Figure 8). Furthermore, interactions between users and $CSPT_m$ within M-Audit are systematically and securely managed. A dedicated phase for auditing is incorporated within the Credential Management Layer of the proposed scheme, as illustrated in Figure 9.

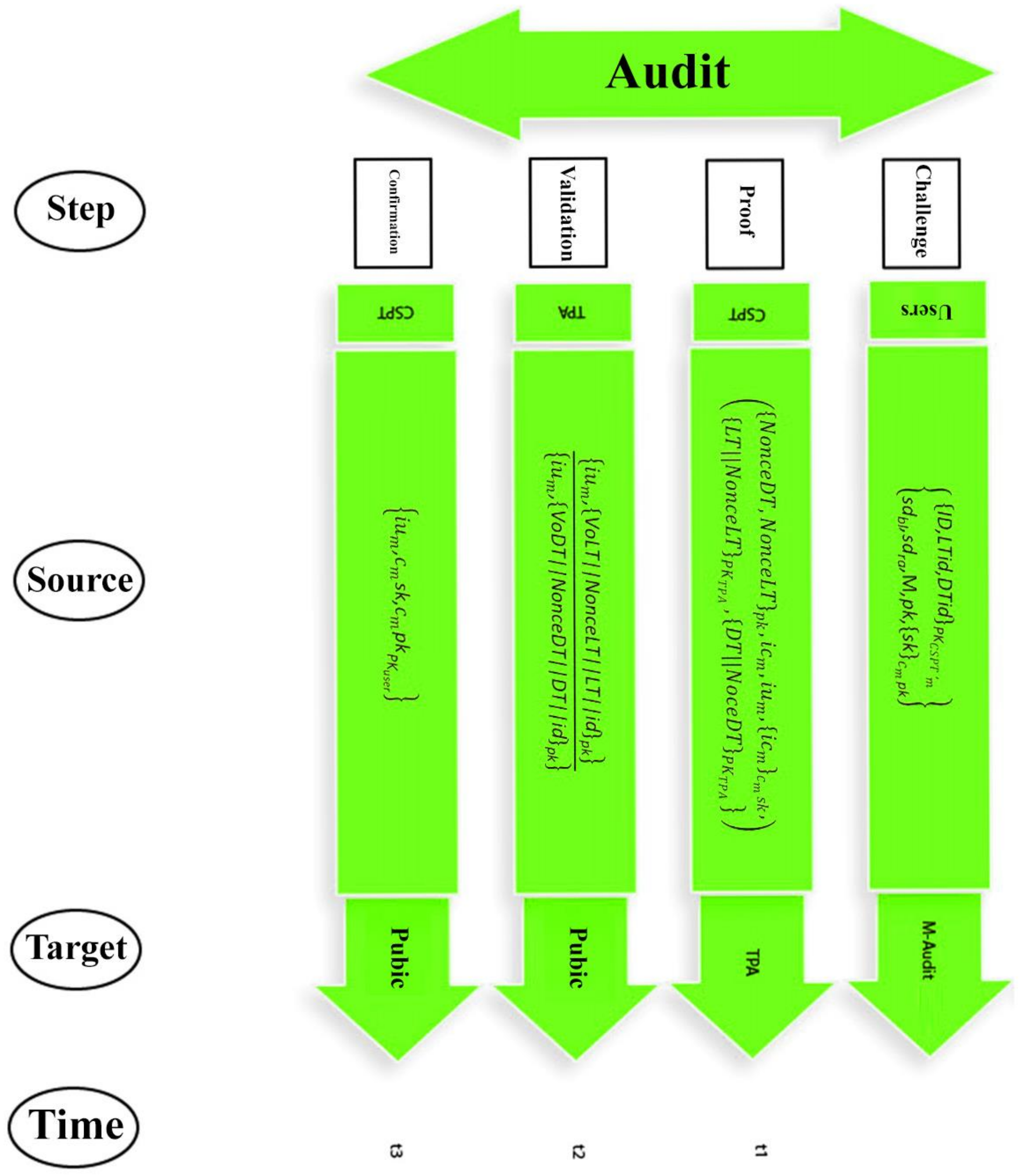

Figure 9. Auditing Process within the Credential Management Layer of the Proposed Scheme

### 3.5. C-AUDIT Layer (Third Layer) in the Proposed Scheme

After completing the encryption stages, attention turns to managing timing and auditing processes. To this end, the C-AUDIT layer is introduced as the third layer of the proposed framework, consisting of two key components:

• Time Coordination,
• Audit Channel Management (M-Audit).

The flowchart in Figure 10 clearly illustrates how the C-AUDIT layer functions within the system.

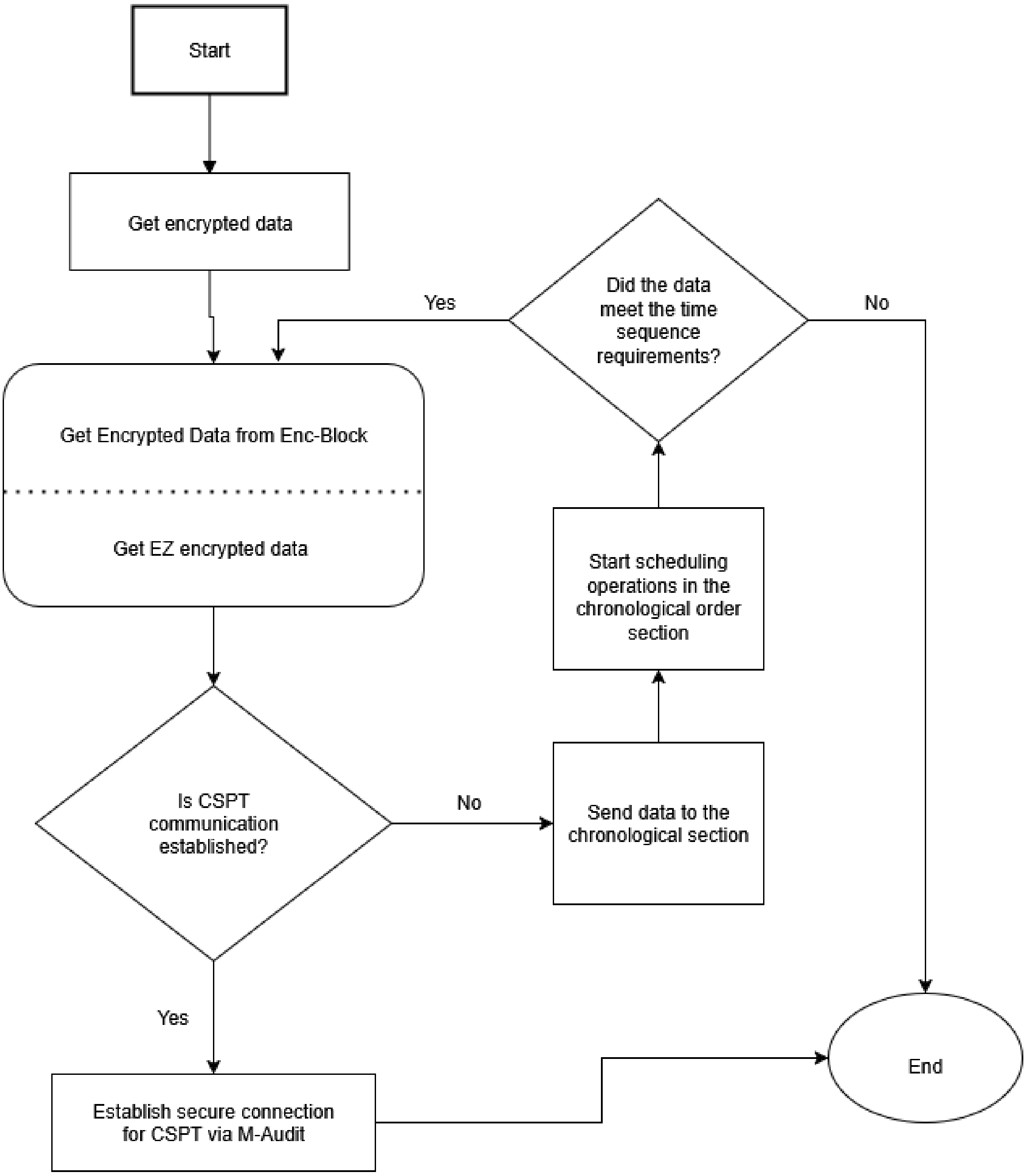

Figure 10. Operational Flowchart of the C-AUDIT Layer

### 3.5.1. Time Coordination in the C-Audit Layer of the Proposed Scheme

To mitigate time-related overhead under conditions of heavy network traffic and increased computational load, the proposed scheme introduces a Time Coordination mechanism within the C-Audit layer. This addition enhances the system's efficiency without undermining its security, as the auditing process remains firmly governed by trusted auditing entities in response to requests initiated by $u_m$.The C-Audit layer systematically organizes tasks in chronological order and dynamically manages timing based on any latency encountered during identity verification. If a response from either $u_m$ or $CSPT_m$ is delayed or entirely absent, C-Audit designates a specific time slot for auditing and data verification, aligned with predefined operational parameters. A visual representation of this Time Coordination mechanism is provided in Figure 11.

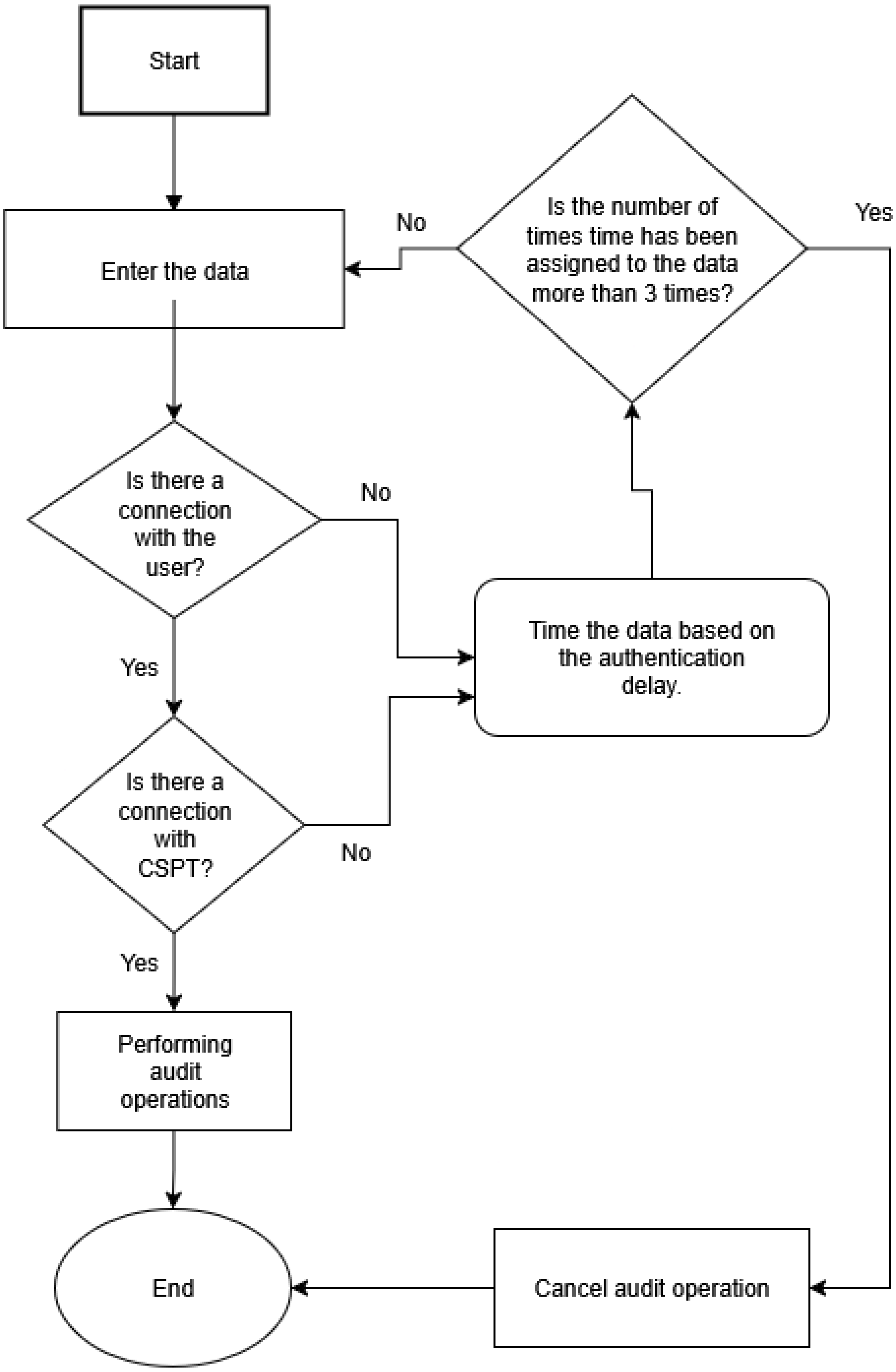

Figure 11. Workflow Diagram of Time Coordination the C-Audit Layer

### 3.5.2. Verification Channel Management (M-Audit) within the C-Verification Tier of the Proposed Framework

The Verification Channel Management unit, referred to as M-Audit, is tasked with supervising all verification-related communication channels established between the various entities involved in the system. Its responsibilities include facilitating inter-entity interactions and managing certain verification tasks specific to each entity.The designation "M-Audit" reflects its central role in orchestrating these exchanges and maintaining a streamlined audit process. Moreover, M-Audit helps to optimize performance by reducing unnecessary communication overhead.

## 4. Analysis of the Suggested Framework

Within this part, an in depth comparative assessment of the suggested framework is conducted, emphasizing four principal aspects: security, time overhead, communication overhead, and the computational cost associated with various operations.

### 4.1. Security Assessment of the Suggested Framework and Comparison with Prior Works

Table 3 presents a comprehensive evaluation of the suggested multi-layer framework against seven security criteria, in comparison with relevant previous studies.As indicated in the table, earlier studies have not fully addressed all these security aspects. For example, foundational schemes like HF-Audit and SEPAR ([30] and [31]) fulfill five of the criteria but lack multi-layer security and data preservation in environments with multiple replicas—both of which are strengths of our proposed approach. Other existing schemes not only miss these features but also differ from our proposal in at least two additional security factors. Overall, the comparison clearly shows that our scheme offers more comprehensive security coverage and outperforms previous research in meeting key security requirements.

Table 3. Comparative Analysis of Security Parameters Across Different Schemes Including the Proposed One

| Attribut / Source | Information-retaining | Consistency across multiple replicas | Verifiable by Third Parties | Accountability Tracking | Distributed | Conspiracy-proof | Layered protection |
|---|---|---|---|---|---|---|---|
| Lu and colleagues (2020) [30] | ✓ | - | ✓ | ✓ | ✓ | ✓ | × |
| Vahi & jassbi (2020) [31] | ✓ | - | ✓ | ✓ | ✓ | ✓ | × |
| Fu and colleagues (2017) [28] | ✓ | - | ✓ | ✓ | × | × | × |
| Huang and colleagues (2014) [26] | × | - | ✓ | × | ✓ | × | × |
| Yu and colleagues (2018) [40] | × | - | ✓ | ✓ | ✓ | × | × |
| Bian and colleagues (2022) [34] | ✓ | - | × | × | × | × | × |
| Suggested Approach | ✓ | ✓ | ✓ | ✓ | ✓ | ✓ | ✓ |

Table 4 provides a review of recent key studies related to the outcomes of the proposed scheme, focusing on a comparison of their security features. A careful analysis of this table shows that our proposed scheme, developed to fill the gaps present in prior research, offers a notably more comprehensive solution than existing approaches. Specifically, the HF-Audit and SEPAR schemes discussed in earlier works ([30] and [31]) encounter several issues, including lack of mutual authentication, susceptibility to physical attacks, anonymity concerns, absence of formal security proofs, and challenges in preventing device identity forgery. When compared side by side, it becomes clear that our scheme effectively overcomes many of these limitations. By enhancing the HF cryptosystems and Enc-Block, integrating EZ, and introducing a dedicated Credential Management Layer alongside time coordination and M-Audit within the C-Audit layer, the proposed approach not only excels beyond HF-Audit and SEPAR across multiple security dimensions but also outperforms other contemporary research efforts presented in Tables 3 and 4. Ultimately, the scheme satisfies all 18 security criteria detailed in these comparisons.

Table 4. Comparative Security Analysis of the Proposed Scheme Against Prior Research

| Source / Attribute | Vahi & jassbi (2020) [31] | Lu and colleagues (2020) [30] | Alshehri & bamasag (2022) [33] | Perera and colleagues (2022) [35] | Fu and colleagues (2017) [28] | Huang and colleagues (2014) [26] | Yu and colleagues (2018) [40] | Bian and colleagues (2022) [34] | Ge and colleagues (2022) [32] | Li and colleagues (2016) [27] | Luo and colleagues (2018) [29] | Suggested Approach |
|---|---|---|---|---|---|---|---|---|---|---|---|---|
| message retransmission attack | ✓ | ✓ | ✓ | ✓ | ✓ | × | × | ✓ | ✓ | ✓ | ✓ | ✓ |
| middle-party intrusion | ✓ | ✓ | ✓ | ✓ | ✓ | ✓ | ✓ | ✓ | ✓ | ✓ | ✓ | ✓ |
| Reciprocal verification | ✓ | × | ✓ | ✓ | ✓ | ✓ | ✓ | ✓ | × | ✓ | ✓ | ✓ |
| Key establishment | ✓ | ✓ | ✓ | ✓ | × | ✓ | ✓ | ✓ | ✓ | × | ✓ | ✓ |
| Fake identity device attack | × | ✓ | ✓ | ✓ | × | ✓ | × | ✓ | ✓ | ✓ | × | ✓ |
| Rogue device insertion attack | ✓ | ✓ | ✓ | ✓ | ✓ | × | ✓ | ✓ | ✓ | ✓ | ✓ | ✓ |
| Physical seizure of device attack | ✓ | × | ✓ | × | × | × | × | × | × | × | × | ✓ |
| Security validation via the AVISPA tool | ✓ | ✓ | ✓ | ✓ | ✓ | × | × | × | ✓ | ✓ | ✓ | ✓ |
| formal security analysis | × | × | ✓ | ✓ | × | ✓ | ✓ | ✓ | ✓ | × | ✓ | ✓ |

| | | | | | | | | | | | |
|---|---|---|---|---|---|---|---|---|---|---|---|
| Authentication without gateway involvement | × | × | × | × | × | × | ✓ | ✓ | × | × | × | ✓ |
| Anonymity protection | × | × | × | × | × | × | × | × | × | × | × | ✓ |

## 4.2. Comparative Analysis of Time Overhead in the Proposed Scheme

Table 5 details the time overheads of the proposed scheme alongside other existing approaches, expressed in milliseconds. Evidently, the proposed method achieves considerably lower time overhead compared to the schemes listed in the table. This improvement is especially pronounced relative to the HF-Audit and SEPAR schemes reported in earlier studies ([30] and [31]). The improved performance results from refining both HF-Audit and SEPAR frameworks, integrating their optimized versions, and incorporating the improved EZ scheme, alongside the use of time coordination, M-Audit, and the oversight provided by the Credential Management Layer.

Table 5. Comparative Analysis of Time Overhead for the Proposed Scheme and Prior Research

| Source / System Configuration | Vahi & jassbi (2020) [31] | Lu and colleagues (2020) [30] | Alshehri & bamasag (2022) [33] | Perera and colleagues (2022) [35] | Fu and colleagues (2017) [28] | Huang and colleagues (2014) [26] | Yu and colleagues (2018) [40] | Bian and colleagues (2022) [34] | Ge and colleagues (2022) [32] | Luo and colleagues (2018) [29] | Proposed Method |
|---|---|---|---|---|---|---|---|---|---|---|---|
| Total rough cost (millisecond) | 252 | 148 | 164 | 213 | 118 | 294 | 129 | 197 | 141 | 126 | 76 |

## 4.3. Evaluation and Comparison of Communication Overhead for the Proposed Scheme

Table 6 outlines the assessment of communication expenses for the proposed model alongside comparable studies. The analysis reveals that our method reduces communication overhead, highlighting its superior efficiency over previous approaches.

Table 6. Communication Cost Comparison Between the Proposed Scheme and Prior Studies

| Source / System Configuration | Vahi & jassbi (2020) [31] | Lu and colleagues (2020) [30] | Alshehri & bamasag (2022) [33] | Perera and colleagues (2022) [35] | Fu and colleagues (2017) [28] | Yu and colleagues (2018) [40] | Bian and colleagues (2022) [34] | Li and colleagues (2016) [27] | Luo and colleagues (2018) [29] | Proposed Method |
|---|---|---|---|---|---|---|---|---|---|---|
| Communication cost (bits) | 4188 | 2031 | 4228 | 3528 | 4108 | 4714 | 3180 | 4010 | 3488 | 1926 |

## 4.4. Comparative Study of Processing Duration for Different Operations within the Presented Framework

This part evaluates the processing duration for critical operations within the presented framework, comparing the outcomes with the HF-Audit and SEPAR frameworks outlined in references [30] and [31].

### 4.4.1. Processing Duration for Bilinear Pairs Tasks within Inter-Chain Networks

Figure 12 illustrates that the processing duration for bilinear pairs tasks within inter-chain networks. The performance in these environments is strongly influenced by the number of TPAs in the referenced frameworks (Lu et al. [30]; Vahi and Jassbi [31]) as well as in the presented approach. As the TPA count increases, the processing time rises proportionally.However, the proposed scheme, benefiting from multi-layer data encryption, consistently shows lower computation times compared to the reference works and experiences a more gradual increase, highlighting its improved efficiency.

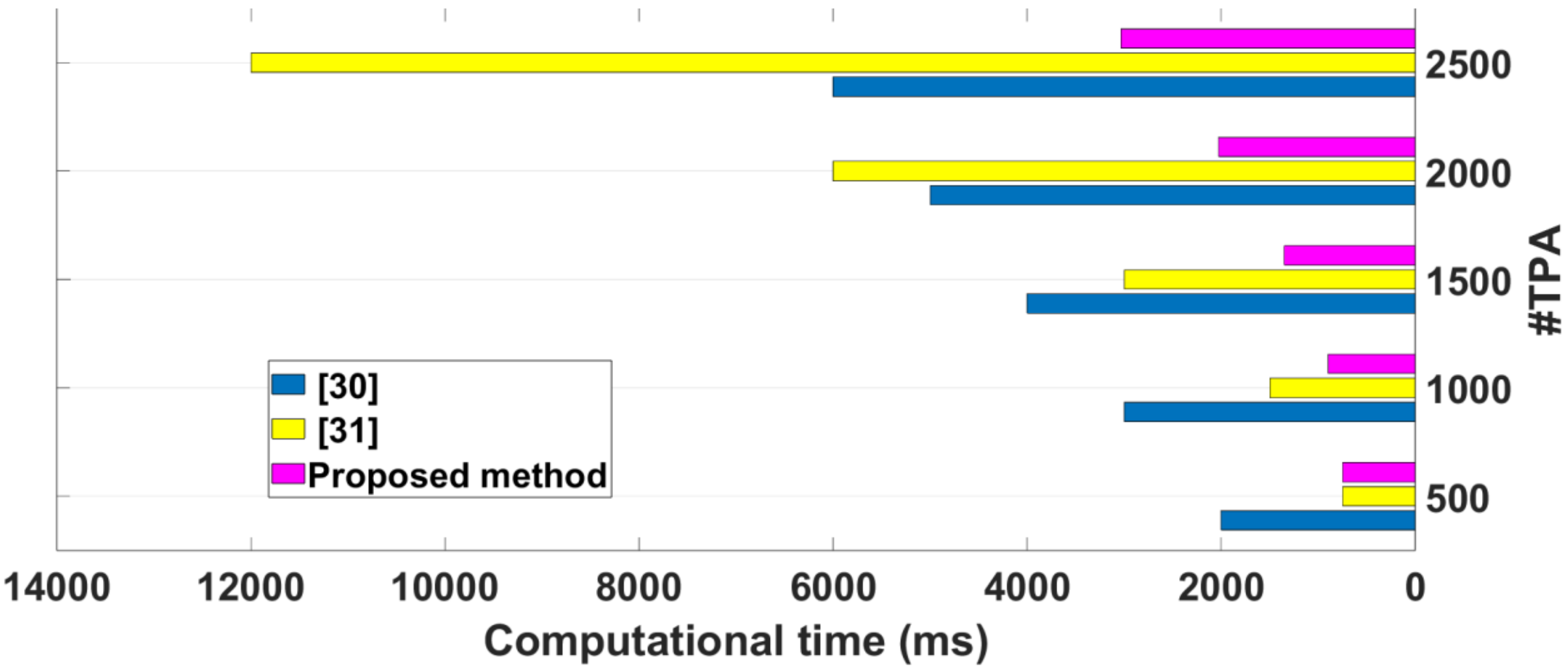

Figure 12. Figure 12. Bilinear Pairs Processing Time in Inter-Chain Networks

### 4.4.2.Processing Duration of Bilinear Pairing Tasks Relative to Block Count within Cross-Chain Networks

Figure 13 shows that a growing count of blocks results in extended bilinear pairing processing times for all evaluated frameworks, encompassing both earlier studies and the presented scheme.For block counts under 100, the growth in computation time is relatively comparable among all three approaches. Yet, as block numbers rise beyond this threshold, the proposed scheme outperforms the others by maintaining better computational efficiency.

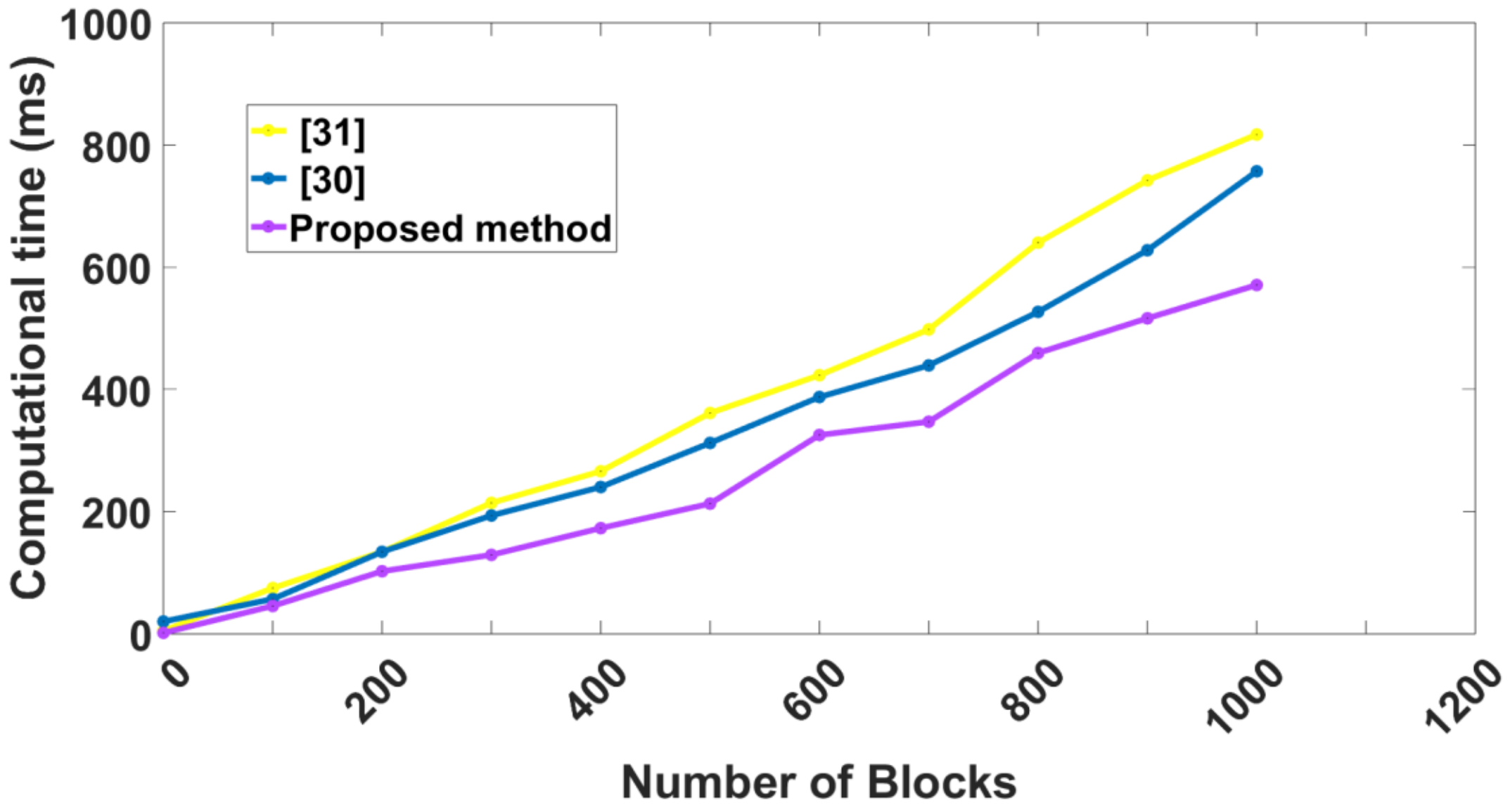

Figure 13. Processing Duration of Bilinear Pairing Tasks Relative to Block Count within Cross-Chain Networks

### 4.4.3 Computation Time in Scenarios with Incomplete Information

Figure 14 shows that as the number of neighboring users increases, the computation time under incomplete information conditions rises for both the proposed scheme and the reference models. Within the presented framework, the minimum processing time of 1100 milliseconds occurs with 30 adjacent participants, rising to a peak of 3100 milliseconds when 120 participants are involved.In comparison, scheme [30] reports a minimum time of 1600 ms and a maximum of 3900 ms for the same user counts. Likewise, scheme [31] exhibits computation times ranging from 1700 ms at 30 neighbors to 4600 ms at 120 neighbors. Overall, the data clearly indicate that the proposed scheme achieves better performance, maintaining lower computation times than the other two approaches.

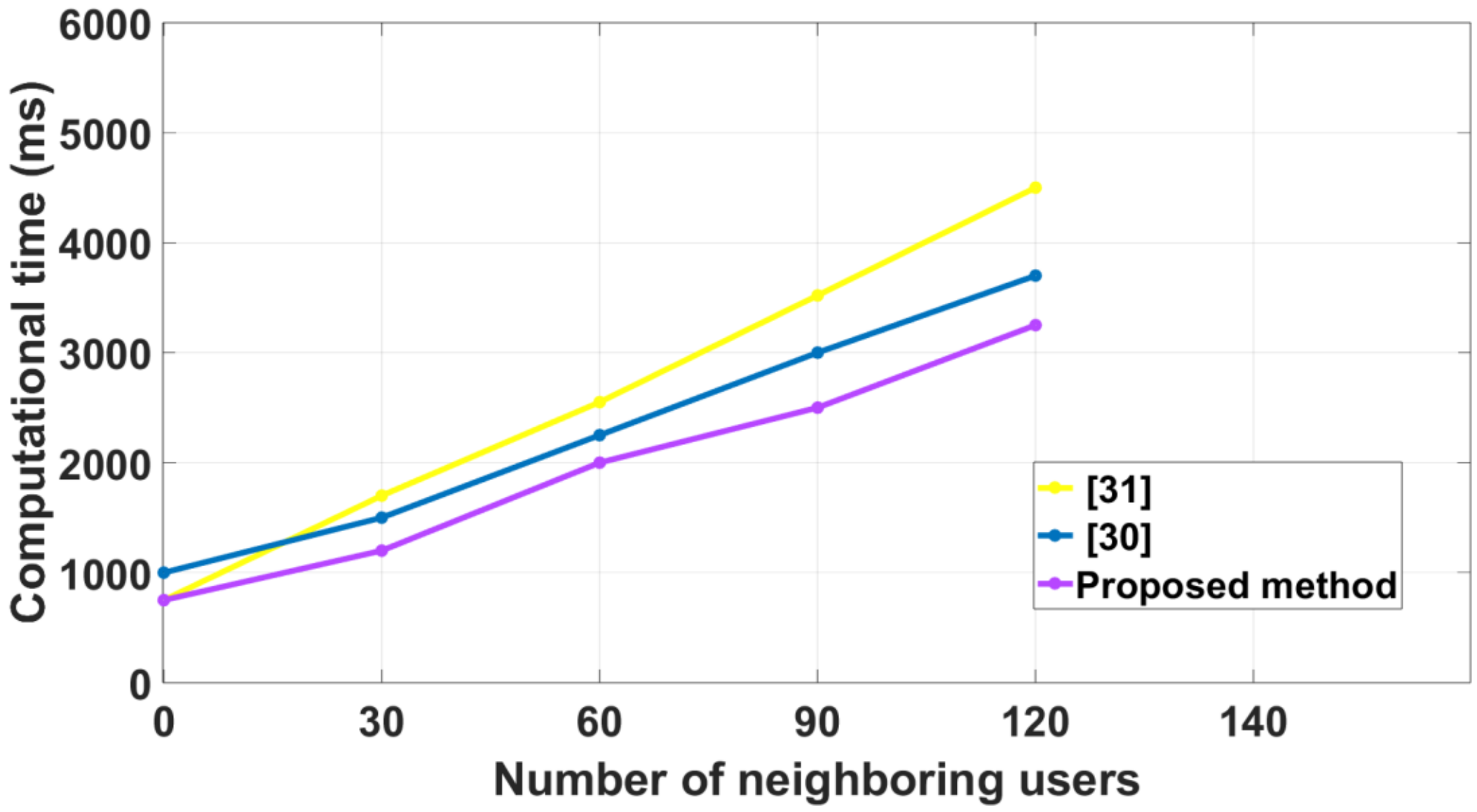

Figure 14. Computation Time under Conditions of Incomplete Information

### 4.4.4.Assessment of Processing Duration for Bilinear Pairing Tasks on the Blockchain

Figure 15 illustrates that the processing duration for bilinear pairing tasks on the blockchain within the presented framework exhibits significantly lower variability and greater stability compared to earlier works.

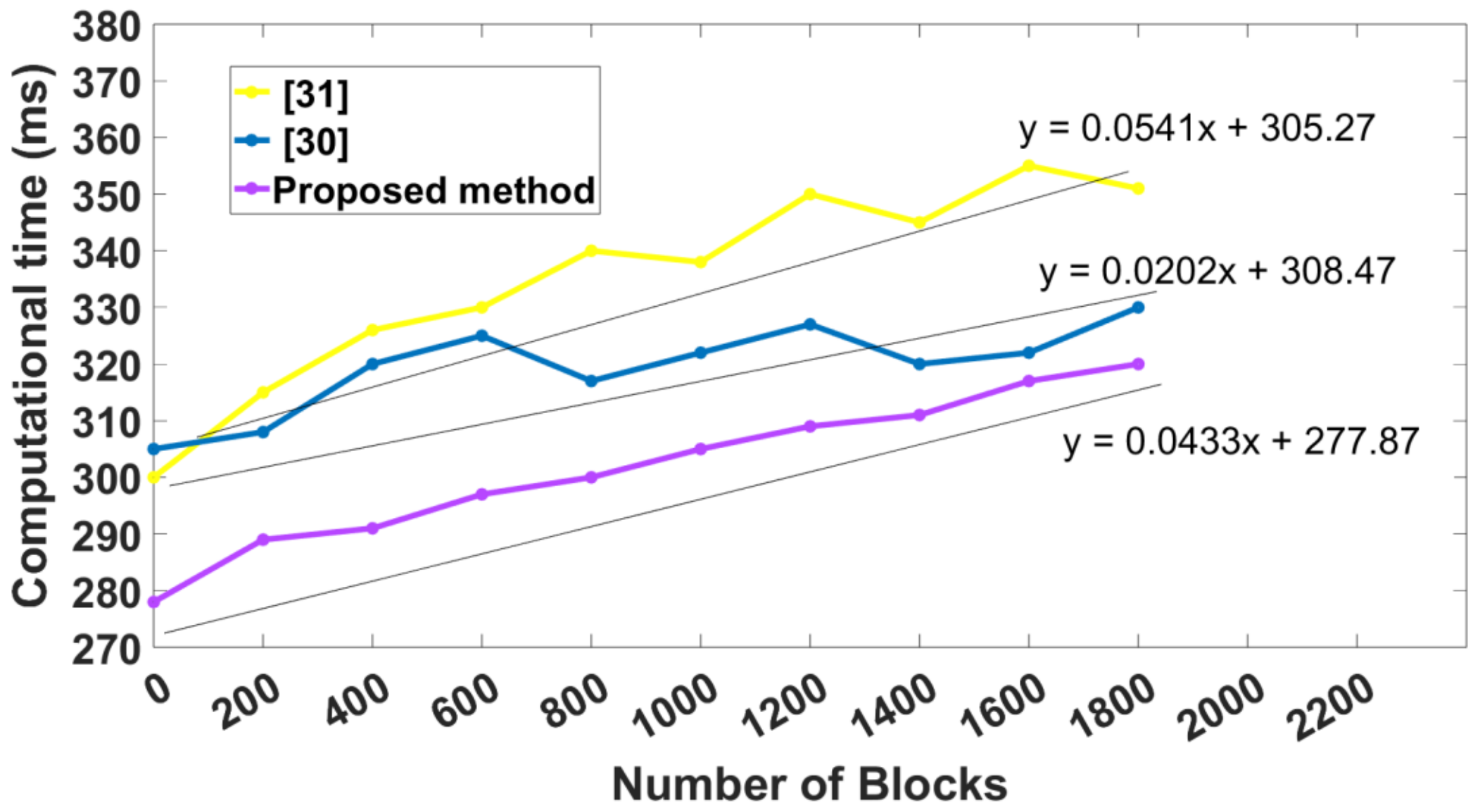

Figure 15. Computation Time of Bilinear Pairing Operation on the Chain

## 5. Conclusion

This research introduced an innovative multi-layered framework aimed at reinforcing data security within IoT-Cloud infrastructures by employing a refined multi-phase lightweight cryptographic design. Faced with the inherent limitations of components such as third-party auditing (TPA), blockchain technology, lightweight encryption methods, and digital signature mechanisms, the proposed scheme was carefully crafted to not only address security vulnerabilities but also to enhance overall system efficiency and processing speed.Our proposed framework, referred to as H.E.EZ in its foundational layer, tackles existing security and performance issues by combining and enhancing three cryptographic methods: Hyperledger Fabric blockchain, Enc-Block and a hybrid ECDSA-ZSS algorithm. This integration improves processing speed and scalability while lowering computational costs and overhead. Beyond this, the scheme boosts data security with its second and third layers—Credential Management and Time & Auditing Management. The Credential Management layer operates independently to validate and ensure the integrity of encrypted data, guaranteeing that information remains reliable throughout every phase. Meanwhile, the Time and Auditing Management layer (C-AUDIT) addresses traffic load, latency, communication overhead and computational timing. It consists of two main components: timestamp ordering and auditing channel management (M-Audit). By scheduling operations based on time and efficiently managing communication channels, this layer significantly enhances the overall system performance. Our findings show that this design not only strengthens data security but also steers the system toward greater sustainability and efficiency by reducing overhead and streamlining processes. In summary, this approach offers a robust and integrated solution for improving security, speed and efficiency in data exchange within IoT-Cloud environments in today's digital age.